\documentclass[%
reprint,
 amsmath,amssymb,
 aps,
]{revtex4-1}

\usepackage{amsmath}
\usepackage{amssymb}
\usepackage{amsfonts}
\usepackage{amssymb}
\usepackage[pdftex]{graphicx}
\usepackage{subfigure} 
\usepackage[english]{babel}
\usepackage{braket}
\usepackage{color}
\usepackage{mathtools}
\usepackage{threeparttable}
\usepackage[colorlinks,linkcolor=blue,anchorcolor=blue,citecolor=blue,urlcolor=blue]{hyperref}
\usepackage{babel}
\makeatother

\begin{document}

\preprint{APS/123-QED}

\title{Controlling evanescent waves using silicon photonic all-dielectric metamaterials for dense integration
}

\author{Saman Jahani${^{1,2}}$}%
\thanks{These authors contributed equally to this work}%
\author{Sangsik Kim${^{2,3}}$}%
\thanks{These authors contributed equally to this work}%
\author{Jonathan Atkinson${^1}$}%
\author{Justin C. Wirth${^2}$}%
\author{Farid Kalhor${^{1,2}}$}%
\author{Abdullah Al Noman${^2}$}%
\author{Ward D. Newman${^{1,2}}$}%
\author{Prashant Shekhar${^1}$}%
\author{Kyunghun Han${^2}$}%
\author{Vien Van${^1}$}%
\author{Raymond G. DeCorby${^1}$}%
\author{Lukas Chrostowski${^4}$}%
\author{Minghao Qi${^{2,5}}$}%
 \email{mqi@purdue.edu}
\author{Zubin Jacob${^{1,2}}$}%
 \email{zjacob@purdue.edu}

\affiliation{%
$^1$Department of Electrical and Computer Engineering,\\University of Alberta, Edmonton, AB T6G 1H9 Canada.
}%
\affiliation{%
$^2$School of Electrical and Computer Engineering and Birck Nanotechnology Center, Purdue University, West Lafayette, IN 47907 USA.
}%
\affiliation{%
$^3$Department of Electrical and Computer Engineering, Texas Tech University, Lubbock, TX 79409
}%
\affiliation{%
$^4$Department of Electrical and Computer Engineering,\\University of British Columbia, Vancouver, BC, V6T 1Z4 Canada.
}%
\affiliation{%
$^5$Shanghai Institute of Microsystem and Information Technology, \\Chinese Academy of Sciences, Shanghai 200050, China.
}%


\begin{abstract}

Ultra-compact, densely integrated optical components manufactured on a CMOS-foundry platform are highly desirable for optical information processing and electronic-photonic co-integration. However, the large spatial extent of evanescent waves arising from nanoscale confinement, ubiquitous in silicon photonic devices, causes significant cross-talk and scattering loss. Here, we demonstrate that anisotropic all-dielectric metamaterials open a new degree of freedom in total internal reflection to shorten the decay length of evanescent waves. We experimentally show the reduction of cross-talk by greater than 30 times and the bending loss by greater than 3 times in densely integrated, ultra-compact photonic circuit blocks. Our prototype all-dielectric metamaterial-waveguide achieves a low propagation loss of approximately $3.67 \pm1$~dB/cm, comparable to those of silicon strip waveguides. Our approach marks a departure from interference-based confinement as in photonic crystals or slot waveguides, which utilize nanoscale field enhancement. Its ability to suppress evanescent waves without substantially increasing the propagation loss shall pave the way for all-dielectric metamaterial-based dense integration.
\end{abstract}

\pacs{Valid PACS appear here}
\maketitle



\section*{Introduction}

One of the long-standing goals of nanophotonics is the integration of electronic and photonic circuitry on a single CMOS chip for applications ranging from information processing and data centers to massively parallel sensing \cite{jalali_silicon_2006, bogaerts_nanophotonic_2005, nagarajan_large-scale_2005, soref_past_2006, kim_dispersion_2017, chrostowski_silicon_2015, dai_passive_2012, heck_hybrid_2013, luo_wdm-compatible_2014, momeni_silicon_2009, shen_deep_2017}. This necessarily requires miniaturization with low power consumption in optical interconnects, active as well as passive photonic devices. There are two major figures of merit in designing photonic devices for a densely integrated circuit. One is the cross-talk, which occurs due to the field overlap of two adjacent photonic waveguides, and the second is the radiation loss at sharp bends which limits the integration density \cite{chrostowski_silicon_2015}.

\begin{figure}[ht]
\centering
\begin{tabular}{cc}
\includegraphics[width=8cm]{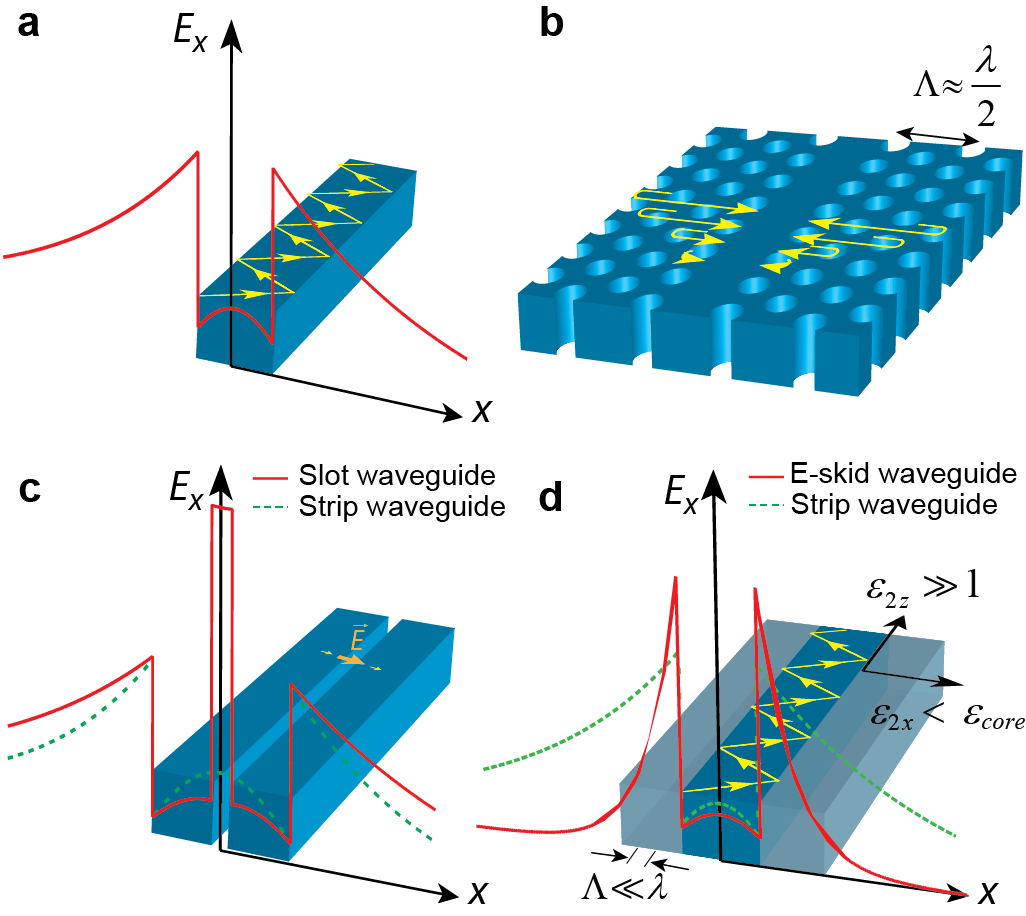}

\end{tabular}
\caption{{\bf Fundamental differences between dielectric waveguides on an SOI platform}. {\bf a,} strip waveguide; {\bf b,} photonic
crystal waveguide; {\bf c,} slot waveguide; {\bf d,} e-skid waveguide. The skin-depth in the cladding of e-skid waveguide is shorter compared to the other structures due to the strong effective anisotropy of the multilayer cladding. This counterintuitive approach can have a cladding with a higher average index than the core and marks a departure from interference based confinement as in photonic crystal waveguides or slot waveguides which utilize nanoscale field enhancement.}
\label{fig:Fig_schematic}
\end{figure}

Plasmonic waveguides can strongly reduce cross-talk and bending loss due to the sub-diffraction nature of light coupling to the free electrons of metals \cite{dionne_plasmon_2006, oulton_hybrid_2008, gramotnev_plasmonics_2010, alu_all_2009, kim_mode-evolution-based_2015, kim_polarization_2015, raza_extremely_2014, kinsey_examining_2015}. However, the large ohmic loss of metals restricts the application of plasmonic structures for photonic integration \cite{khurgin_how_2015, khurgin_replacing_2017}. 
Over the last decade, many efforts have been made to miniaturize photonic components using all-dielectric structures \cite{jahani_all-dielectric_2016, priolo_silicon_2014, kuznetsov_optically_2016, staude_metamaterial-inspired_2017, chang-hasnain_high-contrast_2012, krasnok_all-dielectric_2012, baranov2017all, liu_$q$-factor_2016, arbabi_dielectric_2015, arbabi_multiwavelength_2016, khorasaninejad_metalenses_2016, shen_integrated-nanophotonics_2015, piggott_inverse_2015, hafezi_imaging_2013, slobozhanyuk_three-dimensional_2016}.
Figure~\ref{fig:Fig_schematic} illustrates a few classes of dielectric waveguides for light confinement in photonic chips. Strip waveguides, the most common type of waveguides for routing light in a silicon chip, are composed of a silicon channel surrounded by silicon oxide \cite{chrostowski_silicon_2015} (Fig.~\ref{fig:Fig_schematic}a). Due to the high contrast between the refractive index of the core and the cladding, light is confined inside the core as a result of total internal reflection. However, the mode size is seen to increase as we reduce the core size which hampers the use of strip waveguides to further miniaturize photonic circuits \cite{jahani_transparent_2014}. Photonic crystal waveguides can confine light inside a line defect due to Bragg reflection \cite{joannopoulos_photonic_2008, benisty_optical_1999, hsu_observation_2013} (Fig.~\ref{fig:Fig_schematic}b). These waveguides perform efficiently at very sharp bends \cite{mekis_high_1996}, however, the integration density is limited as the periodicity of Bragg reflectors is on the order of the wavelength and it cannot be perturbed by another waveguide nearby \cite{dai_comparative_2007}. Additionally, slot waveguides have been proposed to confine light inside a sub-wavelength low index gap surrounded by high index dielectric rods \cite{almeida_guiding_2004} (Fig.~\ref{fig:Fig_schematic}c). To satisfy the continuity of the normal component of the displacement current at the high contrast interface, the electric field peaks inside the gap, leading to light confinement but at the cost of skin-depth expansion in the cladding. This causes cross-talk between adjacent waveguides and radiation loss at sharp bends in dense photonic integrated circuits. 

Several alternative approaches inspired from atomic physics \cite{song_high-density_2015, mrejen_adiabatic_2015}, transformation optics \cite{gabrielli_-chip_2012}, and inverse-design algorithms \cite{shen_metamaterial-waveguide_2015, shen_increasing_2016} have also been proposed to minimize the cross-talk and the bending loss. However, these techniques add design complexity often required modification of the core as well as cladding \cite{gabrielli_-chip_2012}. They are not implementable on a large scale and cause propagation loss fundamentally limiting device performance.
Thus, a new low-loss and scalable platform is needed for CMOS foundry compatible dense photonic integration with low cross-talk and reduced bending loss.

In this paper, we demonstrate a platform which is fundamentally different from the existing approaches for designing CMOS compatible, ultra-compact, and low loss waveguides using all-dielectric anisotropic metamaterials. Our approach works based on photonic skin-depth engineering of evanescent waves in the cladding using a recently proposed degree of freedom in total internal reflection (TIR) \cite{jahani_transparent_2014, jahani_photonic_2015}. To describe the light confinement mechanism in our waveguides, first, we experimentally demonstrate the phenomenon of relaxed-total internal reflection in anisotropic metamaterials. These relaxed conditions allow the control of evanescent wave decay which is the fundamental origin of cross-talk and bending loss in silicon photonic devices. As illustrated in Fig.~\ref{fig:Fig_schematic}d, we use these anisotropic metamaterials as a cladding for on-chip dielectric waveguides fabricated on a monolithic silicon-on-insulator (SOI) platform. As a result, cross-talk is reduced down to -30 dB in the photonic circuit. Furthermore, we use a transformation optics approach to show that the radiation loss at sharp bends is strongly influenced by the skin-depth in the cladding. We experimentally show that the anisotropic metamaterial cladding can simultaneously reduce the bending loss at sharp bends up to 3 times compared to conventional silicon strip waveguides. We clarify the counter-intuitive nature of light confinement in our approach compared to existing photonic crystal, slot waveguide, and graded index waveguide methods \cite{joannopoulos_photonic_2008, almeida_guiding_2004, bock_subwavelength_2010, halir_waveguide_2015, yang_giant_2008, levy_implementation_2005}. Our work shows that all-dielectric anisotropy on-chip presents a scalable route to simultaneously improve cross-talk and bending loss with propagation loss as low as $\approx3.67$~dB/cm. For completeness, we show the improvements in figures of merit of our achieved platform with recent state-of-the-art photonic designs (See Table~\ref{crosstalk}).

\begin{figure*}[htbp]
\centering
\begin{tabular}{cc}
\includegraphics{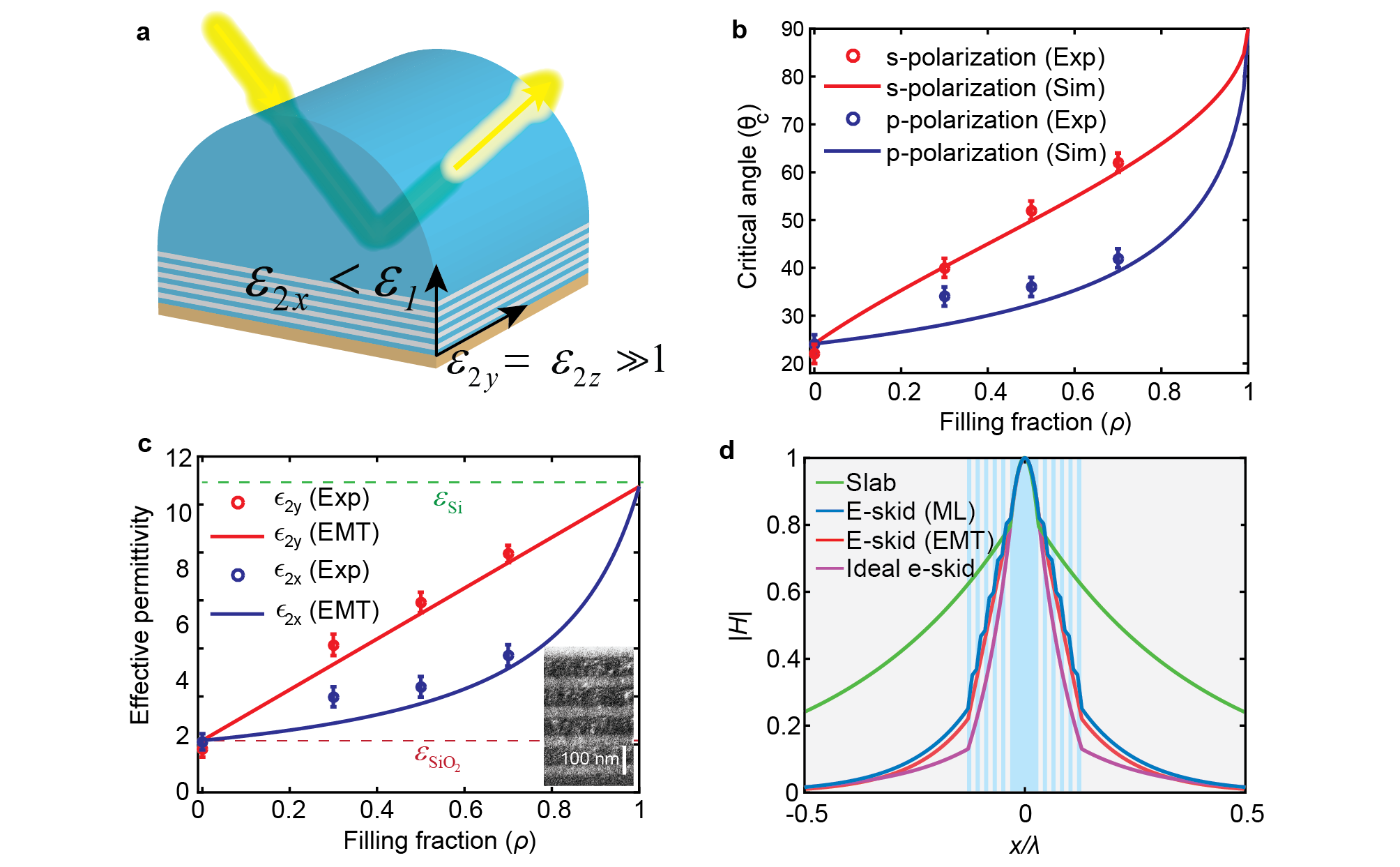}

\end{tabular}
\caption{ {\bf Relaxed total internal reflection}. {\bf a,} Schematic of the relaxed-TIR experiment. We measure the reflection from the Si prism and Si/SiO${}_{2}$ multilayer interface for both {\it s} and {\it p} polarizations at $\lambda=1550$~nm. Multiple samples with different $\rho$ have been fabricated. The periodicity for all samples is 100 nm and the total thickness of the multilayer is 500 nm. A 200 nm layer of tungsten is deposited on top of the multilayer (brown) to suppress the reflection from the air interface. {\bf b,} The measured critical angle for {\it s} and {\it p} polarizations versus the silicon filling fraction in comparison with effective medium theory (EMT) calculations. Error bars represent the instrument limit of the measuring device. In contrast with the conventional phenomenon of TIR, observable differences between the critical angles is seen for different polarizations in agreement with relaxed-TIR theory. {\bf c,} Retrieved effective permittivity of the multilayer from the critical angle measurements show strong anisotropy in agreement with EMT. The refractive index of Si and SiO${}_{2}$ are taken to be 3.4 and 1.47, respectively, for the theoretical calculations. The inset shows the SEM image of an Si/SiO${}_{2}$ multilayer with $\rho=0.5$. {\bf d,} Normalized calculated magnetic field profile for the TM mode of a conventional slab waveguide compared to extreme skin-depth (e-skid) waveguides with Si/SiO${}_{2}$ multilayer cladding. The blue and grey regions represent Si and SiO$_2$, respectively. The core size, $\Lambda$, and $\rho$ are 100 nm, 30 nm, and 0.5, respectively. The multilayer anisotropic claddings strongly affect the decay of the evanescent wave in the cladding. The practical multilayer structure performs close to an ideal anisotropic case ($\varepsilon_{2x}=\varepsilon_{\rm SiO_2}$ and $\varepsilon_{2z}=\varepsilon_{\rm Si}$).}
\label{fig:RTIR}
\end{figure*}

\begin{table}
\centering
\caption{{\bf Performance comparison between an e-skid waveguide and other dielectric waveguides.}}
\label{crosstalk}
\begin{tabular}{|l|c|c|c|}
\hline
\multicolumn{1}{|c|}{\textbf{\shortstack{Reference}}}     & \textbf{\shortstack{Cross-talk}}  & \textbf{\shortstack{Propagation  loss}}\\ \hline
Superlattice \cite{song_high-density_2015}   & -20 dB        & $>20$ dB/cm   \\ \hline
Adiabatic elimination \cite{mrejen_adiabatic_2015}  & -21.9 dB      & N/A           \\ \hline
Inverse design \cite{shen_increasing_2016}     & -22.9 dB      & $>300$ dB/cm  \\ \hline
Dissimilar waveguides \cite{murray_dense_2015}      & -20 dB        & N/A           \\ \hline
Sinusoidal waveguides \cite{zhang_sinusoidal_2015}   & -26.8 dB      & $>600$ dB/cm  \\ \hline
This work (e-skid)                                   & -30 dB        & 3.67 dB/cm     \\ \hline
\end{tabular}
\raggedright
\\
It is seen that the cross-talk is significantly reduced in e-skid waveguides at the negligible cost of propagation loss in comparison with other approaches.
\end{table}

\section*{Results}

{\bf Relaxed total internal reflection.} We demonstrate that only a single component of the dielectric tensor governs total internal reflection (TIR) in anisotropic media opening a new degree of freedom to control evanescent waves.  In conventional TIR,  if $n_{1} >n_{2}$ and the incident angle is greater than the  critical angle ($\theta _{c} =\sin ^{-1} \left({n_{2} \mathord{\left/ {\vphantom {n_{2}  n_{1}}} \right. \kern-\nulldelimiterspace} n_{1} } \right)$), light is reflected back to the first medium and decays evanescently in the second medium. However, we have recently found that the TIR conditions at the interface of an isotropic and an anisotropic dielectric are relaxed to \cite{jahani_transparent_2014, jahani_breakthroughs_2015}:

\begin{align}\label{GrindEQ__1_} 
\begin{array}{ccc} {n_{1} >\sqrt{\varepsilon _{2x}},\quad } & {p \; {\rm polarization} } \end{array} \nonumber \\
\begin{array}{ccc} {n_{1} >\sqrt{\varepsilon _{2y}},\quad } & {s \; {\rm polarization} } \end{array}
\end{align}

where $\left[\begin{array}{ccc} {\varepsilon _{2x} } & {\varepsilon _{2y} } & {\varepsilon _{2z} } \end{array}\right]$ is the permittivity tensor of the second medium, and  the interface between the two media lies on the $yz$ plane. As a result, the critical angle for {\it s} and {\it p} polarizations differ:

\begin{align}\label{GrindEQ__2_} 
\begin{array}{ccc} {\theta _{c} = \sin ^{-1} \left({\sqrt{\varepsilon _{2x} } \mathord{\left/ {\vphantom {\sqrt{\varepsilon _{2x} }  n_{1} }} \right. \kern-\nulldelimiterspace} n_{1} } \right),\quad } & {p \; {\rm polarization} } \end{array} \nonumber \\
\begin{array}{ccc} {\theta _{c} = \sin ^{-1} \left({\sqrt{\varepsilon _{2y} } \mathord{\left/ {\vphantom {\sqrt{\varepsilon _{2x} }  n_{1} }} \right. \kern-\nulldelimiterspace} n_{1} } \right),\quad } & {s \; {\rm polarization} } \end{array}
\end{align}

We emphasize that these relaxed conditions for {\it p} polarized incident light allow us to arbitrarily increase or decrease the permittivity in the $z$ direction while still preserving TIR. This hitherto un-utilized degree of freedom can thus be used to control the skin depth of evanescent waves.  If $\varepsilon _{2z} \gg 1$, evanescent waves decay faster than in vacuum allowing for strong light confinement inside dielectric waveguides \cite{jahani_transparent_2014, jahani_photonic_2015}. In this limit, note that the averaged index in anisotropic medium 2 can be larger than the refractive index in isotropic medium 1 yet the light will be totally reflected above the critical angle.

\begin{figure*}[htbp]
\centering
\begin{tabular}{cc}
\includegraphics{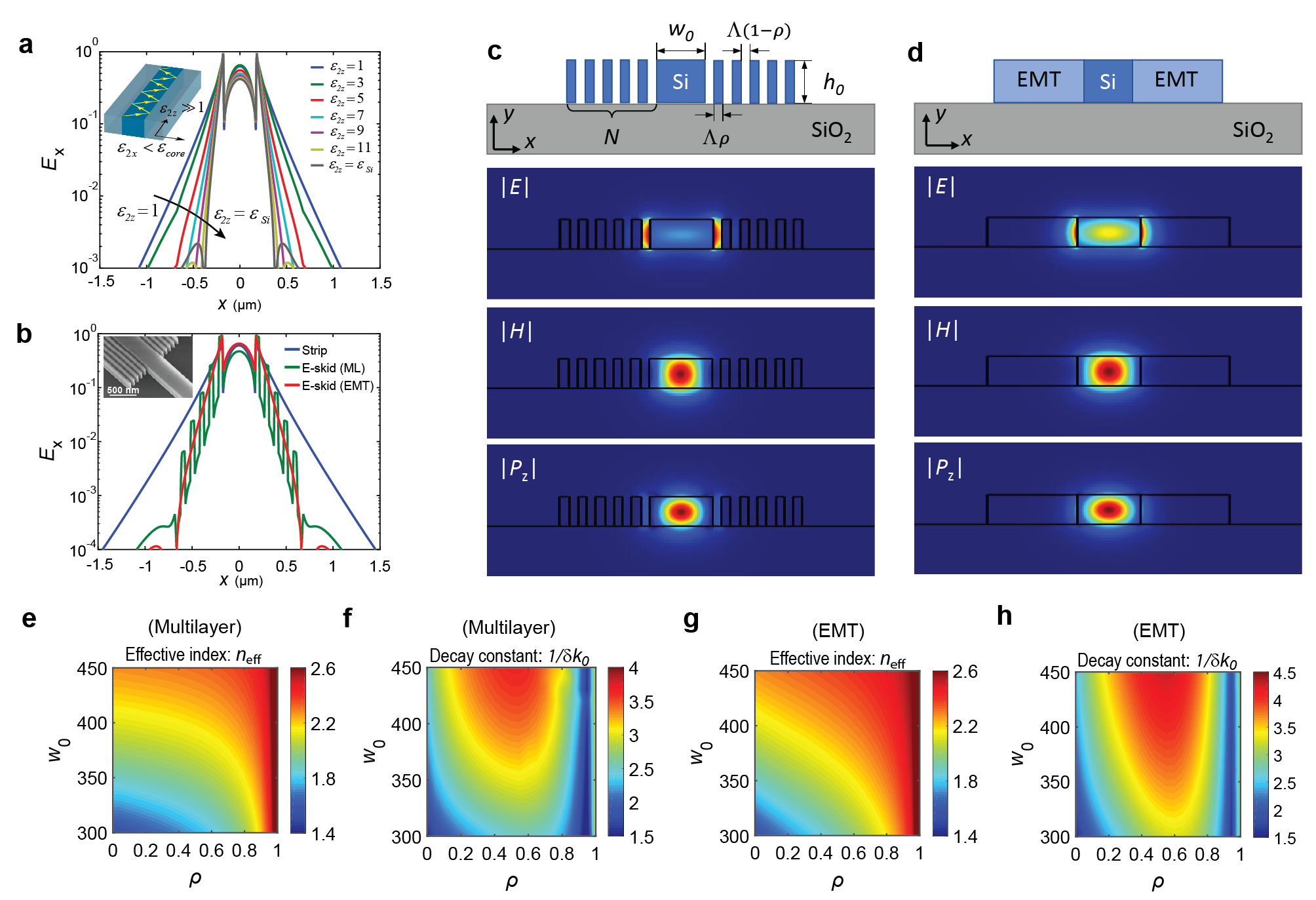}

\end{tabular}
\caption{ {\bf On-chip extreme skin-depth waveguides}. {\bf a,} Ideal on-chip e-skid waveguide; light is confined by total internal refection inside the core and as the effective anisotropy of the multilayer metamaterial cladding is increased, evanescent waves of TE-like modes decay faster in the cladding in comparison with the field in strip waveguides ($\varepsilon _{2z}=1$). Note that $\varepsilon _{2x}=1$ for all cases. {\bf b,} The simulated electric field profile at the center of the e-skid waveguide with multilayer (green) and homogenized metamaterial ($\varepsilon_{2x}=1.85$ and $\varepsilon_{2z}=6.8$) (red) claddings, in comparison with a strip waveguide (blue).  Inset shows the SEM image of the fabricated e-skid to strip waveguide transition. {\bf c, d,} Schematic and field profiles of
({\bf c}) realistic e-skid waveguide with multilayer claddings and ({\bf d}) its equivalent model with EMT claddings.
{\bf e, f,} Effective refractive indices ($n_{\rm eff}=k_z^{\prime}/k_0$) and
{\bf g, h,} normalized decay constants ($k_x^{\prime\prime}/k_0=1/ \delta k_0$)
of the e-skid waveguide as functions of the core width $w_0$ and filling fraction $\rho$:
with ({\bf e, g}) multilayer and ({\bf f, h}) EMT claddings, respectively.
Geometric parameters are $h_0=220$~nm, $w_0=350$~nm, $\Lambda=100$~nm, $\rho=0.5$, and $N=5$, unless otherwise indicated.
The free space wavelength is $\lambda=1550$~nm. Simulations confirm that the cladding achieves effective all-dielectric anisotropy as well as the increased decay constant of evanescent waves outside the core.
}
\label{fig:eskid}
\end{figure*}

\begin{figure*}
\centering
\begin{tabular}{cc}
\includegraphics{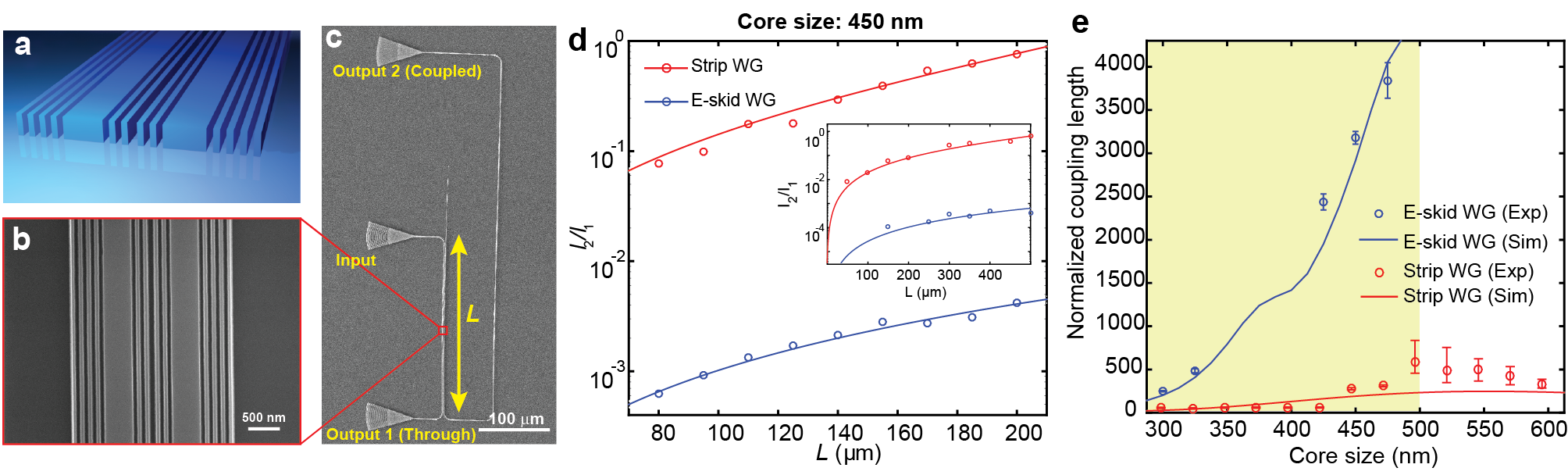}

\end{tabular}
\caption{{\bf Cross-talk in e-skid waveguides}. {\bf a,} Schematic of coupled e-skid waveguides on an SOI platform. The waveguide height, center-to-center separation between the two waveguides ($s$), and $\Lambda$ are 220 nm, 1000 nm, and 120 nm, respectively. A cladding oxide is also added on the top of the waveguides. The number of ridges between the waveguides is dictated by the waveguide core size {\bf b,} Top view SEM image of the coupled e-skid waveguides. {\bf c,} The experimental setup to measure the cross-talk between the two waveguides at the telecommunication wavelength ($\lambda=1550$~nm). Light is in-coupled to the first waveguide through the middle grating coupler. The second waveguide is coupled to the first waveguide for a length of $L$. In this experiment, the bending radius is 5 $\rm\mu$m, hence, we can ignore the bending loss. {\bf d,} The ratio between the measured output powers for strip waveguide and e-skid waveguide versus $L$ at the telecommunication wavelength, respectively. The ratio for the e-skid waveguide is two orders of magnitude lower, indicating that far less power is coupled to the second waveguide. The inset shows the ratio for the waveguides without the top cladding oxide. In this case, the metamaterial cladding can increase the coupling length up to 30 times or reduce the cross-talk -30 dB. See Supplementary Figure 8 for more details. {\bf e,} Comparison of the simulated and measured coupling length for e-skid waveguides and strip waveguides. The coupling length is normalized to the wavelength. Error bars represent the standard deviation of the fitting curves. The optimum match between the simulation and experiment is achieved when $\rho =0.6$ for the cladding of e-skid waveguides. The coupling length for e-skid waveguides is an order of magnitude larger in comparison with strip waveguides (shaded region). The coupling length for strip waveguides with larger core size decreases because the overlap between the evanescent tails is increased although more power is confined inside the core (unshaded region).}
\label{fig:Crosstalk}
\end{figure*}

We implement this anisotropy using Si/SiO${}_{2}$ multilayers with subwavelength thicknesses (Fig.~\ref{fig:RTIR}a) at the telecommunication wavelength ($\lambda=1550$~nm). Figure~\ref{fig:RTIR} highlights the contrast between relaxed-TIR and conventional TIR, showing the measured light reflection at the interface of a hemi-cylindrical Si prism and the anisotropic multilayer metamaterial.  The periodicity of the multilayer is $\Lambda =100$ nm and 5 periods have been deposited on the prism. The Maxwell-Garnett effective medium theory (EMT) \cite{milton_theory_2002} predicts that the multilayer with subwavelength periodicity demonstrates strong anisotropy \cite{jahani_transparent_2014, herzig_sheinfux_subwavelength_2014, halir_ultra-broadband_2016,sayem_broad_2016, gomis-bresco_anisotropy-induced_2017} at the operating wavelength ($\varepsilon _{2y} =\varepsilon _{2z} =\varepsilon _{\rm Si} \rho +\varepsilon _{\rm SiO_{2} } (1-\rho )$  and $1/\varepsilon _{2x} =\rho /\varepsilon _{\rm Si} +(1-\rho )/\varepsilon _{\rm SiO_2 }$ where $\rho$ is the fill fraction of silicon). Note we are in the effective medium metamaterial limit ($\Lambda \ll \lambda$) away from the photonic crystal regime ($\Lambda \sim \lambda$). 
Reflection occurs on two interfaces, the primary silicon-metamaterial interface and the secondary metamaterial-air interface. We have deposited a thin tungsten layer on top of the multilayer metamaterial to attenuate the unwanted reflection from the secondary metamaterial/air interface. 
Figure~\ref{fig:RTIR}b displays different measured critical angles for {\it s} and {\it p} polarizations. Various samples with different fill fractions have been fabricated. As $\rho$ deviates from 0 and 1, the multilayer displays effective anisotropy and the critical angle for {\it s} and {\it p} polarized incidences are separated  in agreement with relaxed-TIR theory in equation \eqref{GrindEQ__2_}. The effective permittivity of the multilayer structure (extracted from the critical angle) shows strong anisotropy and is in good agreement with EMT (Fig.~\ref{fig:RTIR}c). It is clearly seen that the critical angle depends on only one component of the permittivity tensor for each polarization irrespective of the fill-fraction. With the extra degree of freedom afforded by the other components of the dielectric tensor, this anisotropic structure can be used as a cladding for conventional dielectric waveguides to control the evanescent waves and skin-depth in the cladding \cite{jahani_transparent_2014,li_flexible_2016}. Figure~\ref{fig:RTIR}d displays the effect of the multilayer anisotropic cladding on reducing of the evanescent wave skin-depth as compared to a conventional slab waveguide. Note that this evanescent wave engineering approach in extreme skin-depth (e-skid) waveguides is due to the fast variation of the evanescent fields in a subwavelength high-index-contrast super-lattice (Fig.~\ref{fig:RTIR}d). This is fundamentally different from other light confinement strategies which function on interference of propagating waves (photonic crystals) or field enhancement in a slot \cite{joannopoulos_photonic_2008, almeida_guiding_2004, halir_waveguide_2015, yang_giant_2008, levy_implementation_2005}.

{\bf On-chip e-skid waveguides.} To impact the field of silicon photonics, the one-dimensional e-skid waveguide (Fig.~\ref{fig:RTIR}d) has to be implemented on an SOI platform. This places stringent restrictions on monolithic fabrication, minimum feature sizes, and deviations in field profiles due to quasi-2D behavior. Our design to adapt relaxed TIR and skin-depth engineering ($\delta \ll \lambda$)  on-chip is shown in Fig.~\ref{fig:eskid}. Here the conventional quasi-2D strip waveguide is surrounded by a cladding made of alternating vacuum-silicon sub-wavelength ridges which achieves strong anisotropy ($\varepsilon_{2x}<\varepsilon_{\rm core}$ and $\varepsilon_{2z} \gg 1$). Our goal is to confine the TE-like mode (its dominant electric field component is $E_{x}$) used in conventional silicon strip waveguides. Since the TM-like mode is polarized in the $y$ direction, the electric field does not probe/feel the anisotropy of the cladding. Hence, the metamaterial cladding does not play any role in TM-mode confinement. For more details about the polarization effect, see Supplementary Figures 6 and 7.

Figure~\ref{fig:eskid}a emphasizes the counter-intuitive nature of the confinement caused by anisotropic media which beats the confinement of conventional high index contrast interface consisting of silicon surrounded by vacuum  (blue curve in Fig.~\ref{fig:eskid}a). As it is seen in Fig.~\ref{fig:eskid}a, if we increase the index of the cladding only in the $z$ direction while the index in the other directions are fixed, the evanescent wave decays faster in the cladding and the skin-depth reduces drastically. Note, the average cladding index increases but the mode decay length decreases. This behavior is in fundamental contrast to confinement in conventional strip waveguides with isotropic claddings or graded index claddings \cite{levy_implementation_2005} where decreasing the contrast between the core index and cladding index results in slow decay of evanescent waves, not faster decay \cite{ jahani_photonic_2015}.

Figure~\ref{fig:eskid}b shows the field profile of a practical e-skid waveguide with the multilayer metamaterial cladding ($\Lambda \ll \lambda$). We can see the practical multilayer cladding shows strong effective anisotropy to ensure the fast decay of the evanescent wave in the cladding. The 2D field profiles in a plane perpendicular to the propagation direction and comparison of the practical e-skid waveguide with a homogeneous anisotropic cladding are shown in Fig.~\ref{fig:eskid}c and \ref{fig:eskid}d.  A  good agreement with the effective medium approximation is seen. Figure~\ref{fig:eskid}e-\ref{fig:eskid}h illustrate the effective modal index and the decay constant in the cladding for a wide range of core sizes and the fill fractions of silicon in the multilayer cladding. First, a strong agreement is seen between EMT and practical multilayer structures indicating that the cladding indeed achieves effective all-dielectric anisotropy. Secondly, the effective modal index is below the core index indicative of relaxed total internal reflection as the mechanism of confinement. As expected, simulations prove that the primary role of the on-chip multi-layer cladding is to increase the decay constant of evanescent waves ($k_x^{\prime\prime}/k_0$), as shown in Figs.~\ref{fig:eskid}f and \ref{fig:eskid}h, and it does not significantly change the effective modal index of propagating waves ($k_z^{\prime}/k_0$), as shown in Figs.~\ref{fig:eskid}e and \ref{fig:eskid}g. It is seen that the largest decay constant is achieved at $\rho\approx0.5$ where the strongest anisotropy is attained. Note that $\rho=0$ corresponds to the case of no cladding. We also emphasize that the power in the core is not compromised by the presence of the skin-depth engineering cladding and increased anisotropy causes an enhancement in power confinement.  

\begin{figure}
\centering
\begin{tabular}{cc}
\includegraphics[width=8cm]{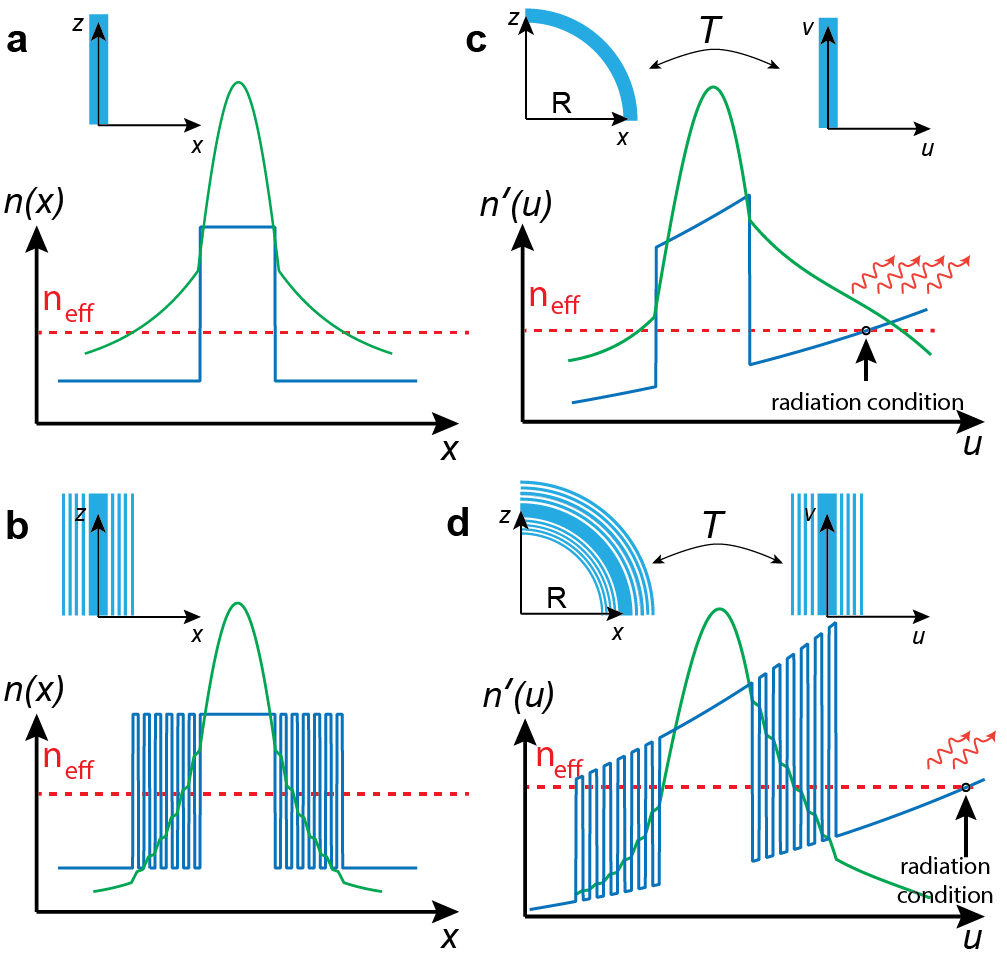}

\end{tabular}
\caption{{\bf Curved waveguides and skin-depth engineering}. {\bf a,} {\bf b,} Refractive index profile (blue) and the magnetic field profile (green) of (a) a straight strip waveguide and (b) a straight e-skid waveguide. {\bf c,}  The corresponding bent waveguides (c,d) can be transformed to straight waveguides with inhomogeneous refractive index profiles in a new coordinate system.  If the local refractive index of the cladding exceeds the effective modal index ($n'(u_0)>n_{\rm eff}$), the waveguide mode starts radiating. This occurs at the radiation condition point denoted by $u=u_0$ (red arrows). The multilayer metamaterial cladding in e-skid waveguides suppresses the evanescent wave field beyond this radiation condition point. Thus, in comparison with conventional bent waveguides, lesser power is radiated due to curvature effects.}
\label{fig:Bending mapping}
\end{figure}

This light confinement strategy in e-skid waveguides is fundamentally different from that in photonic crystal waveguides.  Our multilayer structures function in the deep subwavelength limit  ($\Lambda \ll \lambda$), not the photonic crystal limit ($\Lambda \sim \lambda$) and the performance of the e-skid waveguide is independent of the multilayer periodicity or disorder (see Supplementary note 2). Furthermore, there are no Dyakonov wave solutions in e-skid waveguides as the optical axis of the anisotropic cladding is perpendicular to the interface and the direction of propagation \cite{dyakonov_new_1988, takayama_observation_2009, takayama_lossless_2014, jahani_all-dielectric_2016, polo_surface_2011, talebi_wedge_2016}. It is also important to note that the main goal of the anisotropic multilayer cladding is to control the decay constant of evanescent waves on-chip, not to achieve birefringence in the effective modal index of propagating waves as achieved previously in sub-wavelength grating structures \cite{chang-hasnain_high-contrast_2012, bock_subwavelength_2010, halir_waveguide_2015, boroojerdi_two-period_2016}, multi-slotted waveguides \cite{yang_giant_2008}, or nonlinear phase matching applications \cite{fiore_phase_1998}. 

\begin{figure*}[htbp]
\centering
\begin{tabular}{cc}
\includegraphics{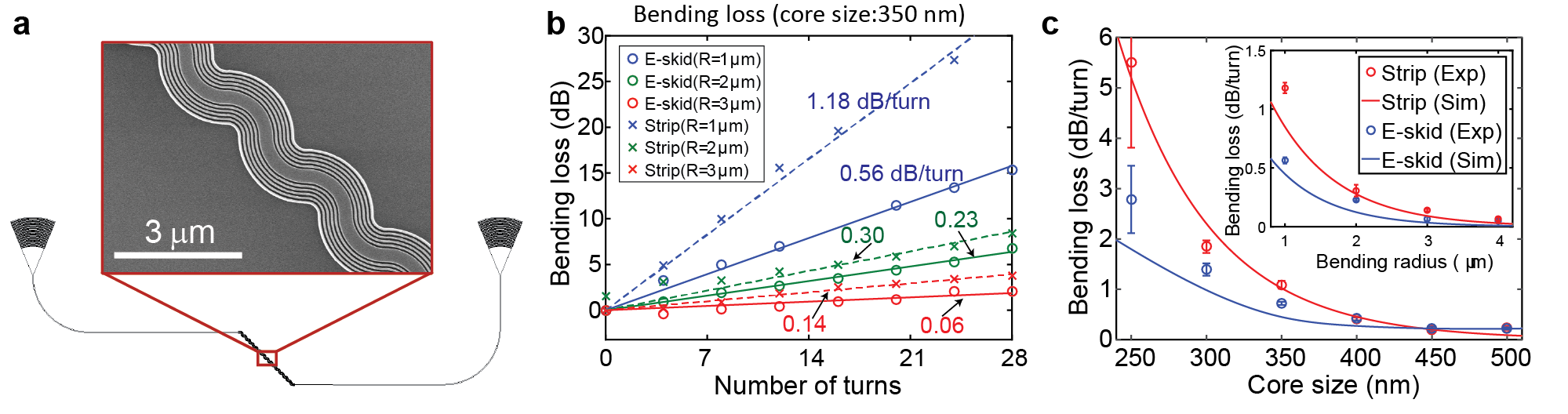}

\end{tabular}
\caption{{\bf Bending loss in e-skid waveguides.}
{\bf a,} Layout and zoomed-in SEM images of the test devices to characterize the bending loss.
{\bf b,} Measured bending losses with different number of turns;
circles and crosses are the e-skid and strip waveguides, respectively,
and blue, green, and red are at different bending radiuses of $R=1$, 2, and 3~$\rm\mu$m.
Each experimental result is fitted with a linear line: e-skid (solid) and strip (dashed) lines.
The core size is $w_0=350$~nm.
{\bf c,} Characterized bending loss vs. core size ($R=1~\rm\mu$m): e-skid (red) and strip (blue) waveguides.
Inset shows the characterized bending loss vs. bending radius ($w_0=350$~nm).
Red and blue lines are the simulated bending losses of the e-skid and strip waveguides, respectively. Error bars represent the standard deviation of the fitting curves.
Other parameters ($h_0, \Lambda, \rho, N$, and $\lambda$) are the same as in Fig.~\ref{fig:eskid}. Confinement of the evanescent waves in e-skid waveguides due to the metamaterial cladding helps to reduce the bending loss at sharp bends in photonic integrated circuits. This effect is stronger for waveguides with smaller core sizes, as a considerable amount of the power lies outside the core and decays slowly.}
\label{fig:Bending loss}
\end{figure*}

{\bf Cross-talk in e-skid waveguides.} The skin-depth of evanescent waves is the fundamental origin of cross-talk between waveguides that hinders dense photonic integration. Cross-talk, or power coupling between photonic devices,  arises due to the perturbation of the optical mode when the evanescent tail of one waveguide overlaps with a nearby waveguide. However, if we control the skin-depth in the cladding, we can reduce the perturbation due to the adjacent waveguide, subsequently reducing the cross-talk to surpass the integration limit in current silicon photonics. 

We now demonstrate this drastic reduction in the cross-talk made possible with the anisotropy in the cladding (Fig.~\ref{fig:Crosstalk}a). The top view image of coupled e-skid waveguides fabricated on an SOI chip is shown in Fig.~\ref{fig:Crosstalk}b. The power exchanged between two identical lossless coupled waveguides is:

\begin{align}\label{GrindEQ__3_} 
I_{1} =I_{0} \cos ^{2} \left(\frac{\pi }{2L_{c} } L\right),\nonumber \\
I_{2} =I_{0} \sin ^{2} \left(\frac{\pi }{2L_{c} } L\right),
\end{align}

where $I_{0} $ is the input power, $L$ is the distance for which the two waveguides are coupled, and $L_{c} $ is the coupling length. Note $L_{c} $ is defined such that if the waveguides are coupled for a distance $L=L_{c}$, complete power transfer occurs from the first waveguide to the adjacent waveguide. Figure~\ref{fig:Crosstalk}c illustrates the experimental setup to measure the coupling length. Light is in-coupled to the first waveguide through the grating coupler \cite{wang_focusing_2014} and the second waveguide is coupled to the first waveguide over a distance of $L$. The center-to-center separation between the two waveguides is only 1000 nm. The output power of the two waveguides is out-coupled by the two grating couplers at the ends of each waveguide. Note that we can ignore the propagation loss in this experiment as the total length does not exceed a few hundred microns whereas the propagation loss is 3.67 dB/cm at the operating wavelength $\lambda=1550$~nm (see Supplementary note 6 for details about the characterization of the propagation loss). The ratio between the measured powers for strip waveguides and e-skid waveguides are shown in Fig.~\ref{fig:Crosstalk}d. This highlights that the analytical expression (${I_{2} \mathord{\left/ {\vphantom {I_{2}  I_{1} =\tan ^{2} }} \right. \kern-\nulldelimiterspace} I_{1} =\tan ^{2} } \left({\pi L\mathord{\left/ {\vphantom {\pi L 2L_{c} }} \right. \kern-\nulldelimiterspace} 2L_{c} } \right)$) matches the experimental data. It is clearly seen that the coupled power to the second waveguide is almost two orders of magnitude lower for the e-skid waveguide. The measured and simulated \cite{CST} coupling length normalized to the wavelength for different core sizes is plotted in Fig.~\ref{fig:Crosstalk}e for waveguides with the same separation distance of 1000 nm. It is seen that the coupling length for the TE-like mode of e-skid waveguides is an order of magnitude higher than that for strip waveguides allowing miniaturization of photonic integrated circuits without considerable cross-talk between the waveguides. Note that increasing the core size in conventional strip waveguides, which corresponds to higher power confinement inside the core, does not guarantee the reduction in cross-talk (unshaded region Fig.~\ref{fig:Crosstalk}e) as the evanescent tails are not controlled.
These drastically increased $L_c$ are due to the reduced skin-depth with a higher anisotropy, reducing the waveguide cross-talk significantly in comparison with other dielectric structures which have been proposed to reduce the cross-talk (Table~\ref{crosstalk}). Note that since the TM-like mode has a negligible electric field component in the $x$ direction, it does not probe/feel the dielectric anisotropy of the cladding. As expected, this results in higher cross-talk for TM-modes compared to strip waveguides (see Supplementary Figures 6 and 7 for more details about the polarization effect and the proposed design to control the skin-depth of the TM-like mode).

{\bf Bending loss in e-skid waveguides.} The critical phenomenon of bending loss \cite{heiblum_analysis_1975, vlasov_losses_2004, marcatili_bends_1969} can also be reduced by adding the anisotropic metamaterial cladding. The counterintuitive connection of bending loss and skin-depth is revealed by an approach adapted from transformation optics \cite{heiblum_analysis_1975, han_calculation_2013}. Here, we consider a curved waveguide in the $xz$ plane with a $90^{\circ}$ bend and a bending radius of $R$ ($R \gg \delta$). The center of the curved waveguide is at the origin (Figs.~\ref{fig:Bending mapping}b and \ref{fig:Bending mapping}d (inset)). If we apply the transformation: $\begin{array}{cc} {u+iw=R\ln \frac{x+iz}{R} ,} & {v=y} \end{array}$ \cite{heiblum_analysis_1975, han_calculation_2013}, the curved waveguide is mapped to a straight waveguide in the $uw$ plane (Figs.~\ref{fig:Bending mapping}b and \ref{fig:Bending mapping}d). This causes the refractive index of the transformed waveguide to become inhomogeneous in the new coordinate system as \cite{heiblum_analysis_1975, han_calculation_2013}: $n'\left(u\right)=n\left(x(u)\right)e^{{u\mathord{\left/ {\vphantom {u R}} \right. \kern-\nulldelimiterspace} R} }$, where $n\left(x\left(u\right)\right)$ is the refractive index of the straight waveguide as shown in Figs.~\ref{fig:Bending mapping}a and \ref{fig:Bending mapping}b. Bending loss or radiative power leakage from the core occurs in this straight inhomogeneous index waveguide when the local index of the cladding exceeds the effective modal index ($n_{\rm eff}$). Hence, if we suppress the field near and beyond this radiation condition point (Fig.~\ref{fig:Bending mapping}b and \ref{fig:Bending mapping}d), we can reduce the bending loss. Note that due to the spatial transformation of coordinates, the electromagnetic fields are also transformed causing the expansion of the skin-depth ($\delta$) on the right-hand side of the waveguide and shrinkage of it on the left-hand side (Figs.~\ref{fig:Bending mapping}b and \ref{fig:Bending mapping}d):

\begin{align}\label{GrindEQ__4_} 
\delta_{\rm{right}} &\approx \delta+2\delta^2/R,\nonumber \\
\delta_{\rm{left}} &\approx \delta-2\delta^2/R.
\end{align}

However, if we add the anisotropic metamaterial cladding, we can reduce the skin-depth. As a result, less power will be radiated at the radiation condition point, leading to reduced bending loss. 

To confirm the effect of the anisotropic metamaterial claddings on the bending loss, we have investigated the bending losses of the e-skid and strip waveguides, both experimentally and numerically \cite{lumerical_fdtd_solutions}. To characterize the bending losses, we have cascaded $90^{\circ}$ bent waveguides sequentially with different numbers of turns (Fig.~\ref{fig:Bending loss}a and Supplementary note 4), then characterized the bending loss per turn by comparing the transmissions. Figures~\ref{fig:Bending loss}b and \ref{fig:Bending loss}c shows the measured bending losses vs. the core size and the bending radius. In all cases, the bending losses with the e-skid waveguide is lower than that with the strip waveguide, which is due to the reduced skin-depth with the anisotropic metamaterial claddings. Note that if the fabrication advancements on the CMOS platform allowed us to reduce the feature size and approach an ideal anisotropic cladding ($\Lambda \rightarrow 0$), we would achieve further reduction of the bending loss for TE-like modes in e-skid waveguides (See Supplementary note 4 for more details as well as the bending loss calculation for TM-like modes). Our simulations account for all sources of bending loss including radiation and mode-mismatch.

\section*{Discussion}

In summary, we have introduced a photonic platform which can add the critical but overlooked functionality of controlling evanescent waves to the CMOS foundry.  We have shown that high index contrast grating structures in the deep sub-wavelength limit can act as an all-dielectric metamaterial cladding for simultaneously achieving total internal reflection and controlling the skin-depth of evanescent waves. The coupling length is improved more than an order of magnitude and the bending loss is improved three times compared to conventional on-chip waveguides with an average propagation loss of 3.67 dB/cm at telecommunication wavelengths. The decreased photonic skin depth regime has not been realized till date but has been attempted in initial studies \cite{khavasi_significant_2016}. This is because anisotropic all-dielectric metamaterial response requires more than one period of the unit cell to control evanescent wave decay \cite{jahani_transparent_2014, jahani_photonic_2015, jahani_breakthroughs_2015}. Although we use Electron Beam Lithography (EBL) as a convenient prototyping technique for the sub-wavelength structures, these devices can in principle be fabricated using deep ultraviolet (DUV) CMOS foundries; specifically, advanced 193 nm immersion lithography technology has been used to fabricate silicon photonic devices with feature sizes down to 50 nm \cite{selvaraja_193nm_2014}. Our work paves the way for all-dielectric metamaterials to enter the practical realm of CMOS foundry photonics to achieve improved photonic integrated circuits. 

\section*{Methods}

Si/SiO$_2$ multilayers were fabricated using magnetron sputtering and magnetron reactive sputtering, respectively.  Silicon and silica were both deposited at a power of 150 W using a pulsed power supply at a frequency of 150 kHz and off time of 0.5 $\rm\mu$s for silicon and 0.8 $\rm\mu$s for silica.  A special substrate holder was built to hold the prisms during deposition.  With the new substrate holder and the large thickness of the prism, the film properties would change as the substrate was much closer to the target, producing films with higher loss.  Reducing the deposition power produced lower loss films.
A 200 nm thick layer of tungsten was deposited at 300 W, 150 kHz and 0.5 $\rm\mu$s on top of the multilayer structure at each fill fraction. 
The prism was illuminated with a 1530 nm narrow line width laser. A broadband linear polarizer placed in the beam path created {\it s} and {\it p} polarized light.  The incident angle was increased in increments of 2$^{\circ}$ from 10$^{\circ}$ to 80$^{\circ}$.  A Newport Optics Optical Power Meter calibrated to a wavelength of 1530 nm was then used to measure the reflected power.  The two reflections at the prism/air interface in the optical path were accounted for when comparing experiment to EMT simulations. 

The on-chip devices for measuring cross-talk in the main text and bending loss in Supplementary Figure 9 were fabricated using a JEOL JBX-6300FS Electron Beam Lithography system \cite{bojko_electron_2011} operated at 100 keV energy, 8 nA beam current, and 500 $\rm\mu$m exposure field size. A silicon-on-insulator wafer (220 nm thick silicon on 3 $\rm\mu$m thick silicon dioxide) has been used. A solvent rinse and hot-plate dehydration baked. Then, hydrogen silsesquioxane resist (HSQ, Dow-Corning XP-1541-006) was spin-coated at 4000 rpm, and then hotplate baked at 80 $^{\circ}$C for 4 minutes. Shape placement by the machine grid, the beam stepping grid, and the spacing between dwell points during the shape writing, were 1 nm, 6 nm, and 6 nm, respectively. An exposure dose of 2800 $\rm\mu$C/cm$^2$ was used. The resist was developed by immersion in 25\% tetramethylammonium hydroxide for 4 minutes, followed by a flowing deionized water rinse for 60 s, an isopropanol rinse for 10 s. Then blown dry with nitrogen. The inductively coupled plasma etching in an Oxford Plasmalab System 100 was used to remove silicon from unexposed areas, with a chlorine gas flow of 20 sccm, ICP power of 800 W, pressure of 12 mT, bias power of 40 W, and a platen temperature of 20 $^{\circ}$C, resulting in a bias voltage of 185 V. During etching, perfluoropolyether vacuum oil was used to mount chips on a 100 mm silicon carrier wafer. Cladding oxide was deposited using plasma enhanced chemical vapor deposition (PECVD) in an Oxford Plasmalab System 100 with nitrous oxide (N$_2$O) flow of 1000.0 sccm, a silane (SiH$_4$) flow of 13.0 sccm, high-purity nitrogen (N$_2$) flow of 500.0 sccm, high-frequency RF power of 120 W, pressure at 1400 mT, and a platen temperature of 350 $^{\circ}$C. Chips rest directly on a silicon carrier wafer during deposition and are buffered by silicon pieces on all sides to aid uniformity.

To characterize the on-chip devices, a custom-built automated test setup was used \cite{chrostowski_silicon_2015}.  An Agilent 81635A optical power sensor was used as the output detector and Agilent 81600B tunable laser as the input source. The wavelength was swept  in 10 pm steps from 1500 to 1600 nm. To maintain the polarization state of the light, a polarization maintaining fiber was used for coupling the TE polarization into the grating couplers \cite{wang_focusing_2014}.  A polarization maintaining fiber array was used to couple light in/out of the chip.

For the bending loss experiment in the main text, the cross-talk experiment in Supplementary Figure 8, the insertion loss experiments in Supplementary Figure 13, and the propagation loss experiment in Supplementary Figure 14, a standard SOI wafer is used as a substrate with 2 $\rm\mu$m buried oxide (BOX) and 220 nm top silicon layer. Diluted hydrogen silsesquioxane (HSQ) with methyl isobutyl ketone (MIBK) was spun on the substrate as a negative-tone electron beam resist layer. The resist layer was exposed by a 100 kV electron-beam lithography system, VB6-UHR (Raith), which is capable of 2 nm beam step resolution. After the development of the resist, the top silicon layer was etched by Cl$_2$/O$_2$ based reactive-ion plasma etching tool (Panasonic P610) to transfer the waveguide pattern from the resist to the silicon layer.

{\bf Data availability.} The data that support the findings of this study are available
from the corresponding authors upon request.

\section*{Acknowledgements}

This work is supported by National Science Foundation (DMR-1654676). M. Qi acknowledges partial in-kind support by the Strategic Priority Fundamental Research Program (category B) from the Chinese Academy of Sciences under grant XDB24020200, and a Key Project in Silicon Photonics from the Shanghai Municipal Government. S. Jahani and Z. Jacob acknowledge the edX UBCx Silicon Photonics Design, Fabrication and Data Analysis course, supported by the Natural Sciences and Engineering Research Council of Canada (NSERC) Silicon Electronic-Photonic Integrated Circuits (SiEPIC) Program. Some of the on-chip devices for measuring the cross-talk were fabricated by Richard Bojko at the University of Washington Washington Nanofabrication Facility. Some of the cross-talk measurements were performed by Zeqin Lu at The University of British Columbia. We acknowledge KLayout and SiEPIC EBeam PDK for the design software.

\section*{Author contributions statement}

S.J., S.K., J.A., and F.K. designed the experiments. S.K., J.A., J.C.W., A.A.N., K.H. performed the experiments. S.J., S.K., J.A., A.A.N., and P.S. performed the characterizations and measurements.
S.J., S.K., and W.D.N. analyzed the data. S.J. and S.K. conducted numerical simulations and prepared the figures. S.J., S.K., and Z.J. wrote the manuscript. All authors discussed the results and commented on the manuscript. Z.J. and M.Q. supervised the project.

\section*{Additional information}

{\bf Competing financial interests:} the Authors declare no Competing Financial or Non-Financial Interests.

\cleardoublepage
\renewcommand{\figurename}{Supplementary Figure}
\renewcommand\thesection{ SUPPLEMENTRY NOTE \arabic{section}}

\begin{widetext}
\begin{center}

\begingroup
    \fontsize{14pt}{12pt}\selectfont
    \bf{Supplementary information: controlling evanescent waves using silicon photonic all-dielectric metamaterials for dense integration}
    
\endgroup
\bigskip
Saman Jahani$^{1,2}$*, Sangsik Kim$^{2,3}$*, Jonathan Atkinson$^{1}$, Justin C. Wirth$^{2}$, Farid Kalhor$^{1,2}$,
Abdullah Al Noman$^{2}$, Ward D. Newman$^{1,2}$, Prashant Shekhar$^{1}$, Kyunghun Han$^{2}$, Vien
Van$^{1}$, Raymond G. DeCorby$^{1}$, Lukas Chrostowski$^{4}$, Minghao Qi$^{{2,5}}$, and Zubin Jacob$^{1,2}$\\
\bigskip
$^{1}$ Department of Electrical and Computer Engineering,\\
University of Alberta, Edmonton, AB T6G 1H9 Canada.\\
$^{2}$ School of Electrical and Computer Engineering and Birck Nanotechnology Center,\\
Purdue University, West Lafayette, IN 47907 USA.\\
$^{3}$ Department of Electrical and Computer Engineering,\\
Texas Tech University, Lubbock, TX 79409 and\\
$^{4}$ Department of Electrical and Computer Engineering,\\
University of British Columbia, Vancouver, BC, V6T 1Z4 Canada.\\
$^{5}$ Shanghai Institute of Microsystem and Information Technology, \\
Chinese Academy of Sciences, Shanghai 200050, China.
\end{center}
\end{widetext}

\begin{figure*}[ht]
\centering
\begin{tabular}{cc}

\includegraphics{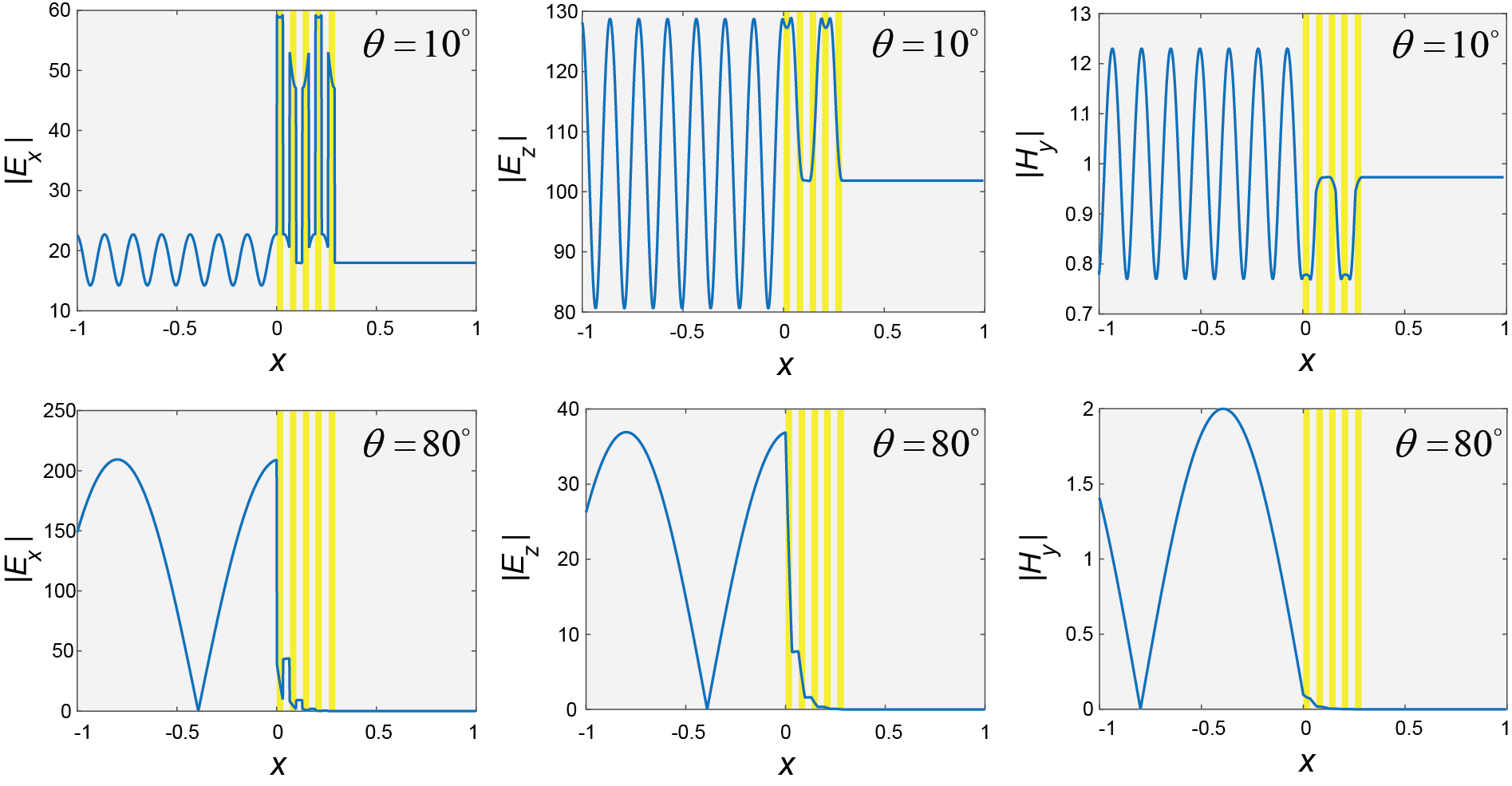}

\end{tabular}
\caption{Comparison of the field profile for the p-polarized electromagnetic wave propagation through an Si/SiO$_2$ multilayer anisotropic  metamaterial below (top) and above (bottom) the critical angle ($\theta_c=32.33^{\circ}$). For both cases, $\Lambda=100$ nm and $\rho=0.5$. It is seen that incident light is totally reflected although the total thickness of the metamaterial layer ($t=500$ nm) is subwavelength ( $\lambda=1550$~nm).}
\label{fig:FigS_TIR_field}
\end{figure*}

\section{\label{sec:level1}Relaxed total internal reflection}

Supplementary Figure~\ref{fig:FigS_TIR_field} displays the field profile for p-polarized plane-wave propagation through an Si/SiO$_2$ multilayer sandwiched between two Si half spaces. The incident angle is $\theta=10^{\circ}$ (top) and $\theta=80^{\circ}$ (bottom). In both cases, $\Lambda=100$ nm and $\rho=0.5$, and light is incident from the left side at $\lambda=1550$ nm. The gray and yellow regions represent Si and SiO$_2$ layers, respectively.  The critical angle of the effective medium is $\theta_c=32.33^{\circ}$  based on relaxed-TIR theory. It is seen that above the critical angle, the electromagnetic wave is totally reflected back to the left side and forms a standing wave on the left side but decays inside the multilayer and negligible power is transmitted to the right half space. However, when the incident angle is below the critical angle, most of the power is transmitted through the multilayer because the total thickness of the slab is thin in comparison with the wavelength.

\section{\label{sec:level1}Effect of periodicity and disorder on e-skid waveguides}

Although practical metamaterial structures for realization of strong effective anisotropy are usually periodic, the effective macroscopic electromagnetic response of metamaterials does not stem from the periodicity (or unit-cell size). As  long as the periodicity of the metamaterial building blocks is deeply subwavelength, the change in the periodicity does not change the electromagnetic response of metamaterials. Indeed, according to the effective medium theory (EMT), which has been explained in the main text, the effective constitutive parameters of multilayer metamaterials depend on the permittivity of the building blocks and their filling fraction, not the periodicity of the multilayer. 

To demonstrate the effect of periodicity on the performance of e-skid waveguides, we analyzed 1D e-skid waveguides with multilayer claddings of the same total size, but different periodicities as illustrated in Supplementary Fig.~\ref{fig:FigS_field_period}. We also analyzed a case where the multilayer is not periodic. We can see that since the effective permittivity for all three cases is the same, the performance of the waveguides are almost identical. This demonstrates that the light confinement mechanism in e-skid waveguides and photonic crystal waveguides are fundamentally different. In photonic crystal waveguides, the periodicity is a requirement and a change in periodicity strongly affects the performance of the waveguide. 

We have also studied the effect of periodicity on the modal effective index ($n_{\rm eff}$) of TE-like and TM-like modes of a single on-chip e-skid waveguide as a function of core size at $\lambda$=1550 nm. Supplementary Figure~\ref{fig:FigS_effective_index} shows that EMT is an extremely accurate approximation to homogenize the cladding when the unit cell size is considerably smaller than the operating wavelength. The effect of the unit-cell size is more obvious on the effective index of the TE-like mode because the field intensity is higher in the cladding region for these modes.

 \begin{figure*}[ht]
\centering
\begin{tabular}{cc}

\includegraphics{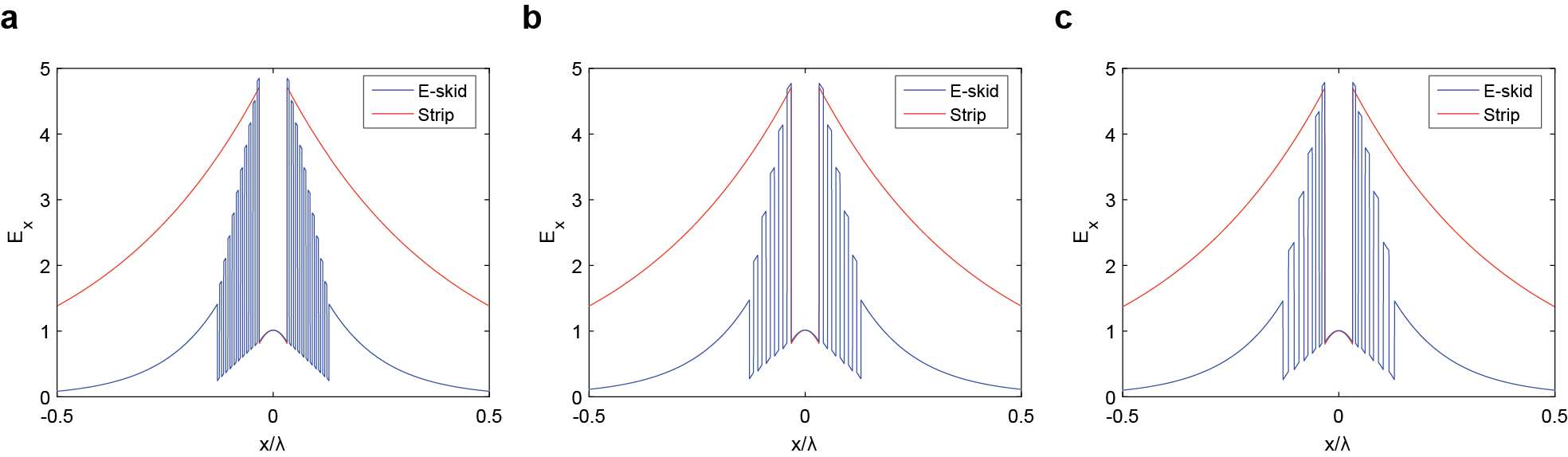}

\end{tabular}
\caption{{\bf Effect of the cladding periodicity on 1D e-skid waveguides.} The comparison of the x-component of the electric field at $\lambda$=1550 nm. The core is silicon with a size of 100 nm. The cladding is Si/SiO$_2$ multilayer with $\rho=0.5$. (a) The periodicity is  $\Lambda$=10 nm. (b) The periodicity is $\Lambda$=30 nm. (c) The periodicity linearly varies from $\Lambda$=20 nm to $\Lambda$=45 nm. The total thickness of the waveguide in all cases is 400 nm. The modal effective index ($n_{\rm eff}$) for the three cases are 1.9062, 1.8864, and 1.8864,
respectively. The ratio between the power confined inside the core and total power ($\eta$) for the three cases are 30.98\%, 29.79\%, and 29.99\%, respectively. $n_{\rm eff}$= 1.5281 and   $\eta$= 12.36\% when the cladding is removed. It is clearly seen that as long as the periodicity is subwavelength, the performance of e-skid waveguide depends on the effective permittivity of the multilayer cladding not the periodicity or disorder in the cladding. Thus the waveguiding mechanism in e-skid waveguides is fundamentally different from that in photonic crystal waveguides.}
\label{fig:FigS_field_period}
\end{figure*}

\begin{figure}[ht]
\centering
\begin{tabular}{cc}

\includegraphics{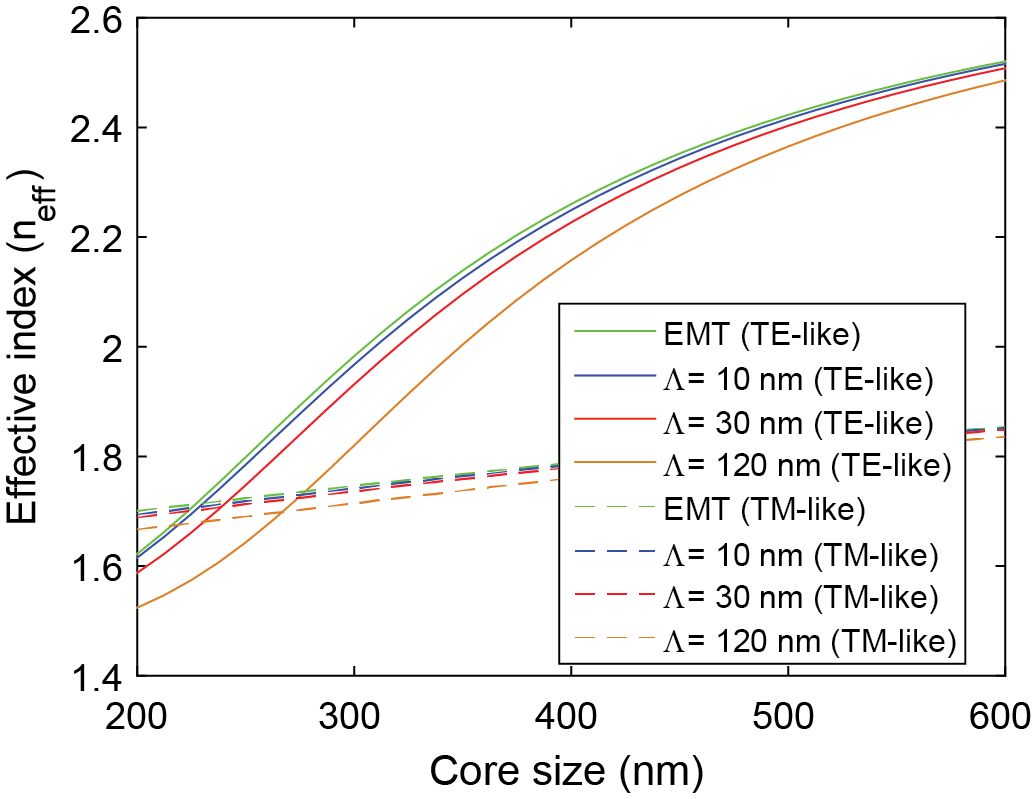}

\end{tabular}
\caption{{\bf Effect of the cladding periodicity on the modal index.} Effect of the multilayer cladding periodicity ($\Lambda$) on the modal effective index of the TE-like and TM-like modes of a single e-skid waveguides on SOI platform. The simulation parameters are the same as those in Fig. 4 in the main text. The effective permittivity of the cladding is  $\left[\begin{array}{ccc} {\varepsilon _{2x} } & {\varepsilon _{2y} } & {\varepsilon _{2z} } \end{array}\right]$ =  $\left[\begin{array}{ccc} {2.33} & {7.6} & {7.6} \end{array}\right]$ which is calculated using effective medium theory (EMT). As $\Lambda \rightarrow 0$, there is an excellent match between EMT and multilayer simulations.}
\label{fig:FigS_effective_index}
\end{figure}

\section{\label{sec:level1}Cross-talk}

\begin{figure*}[ht]
\centering
\begin{tabular}{cc}

\includegraphics{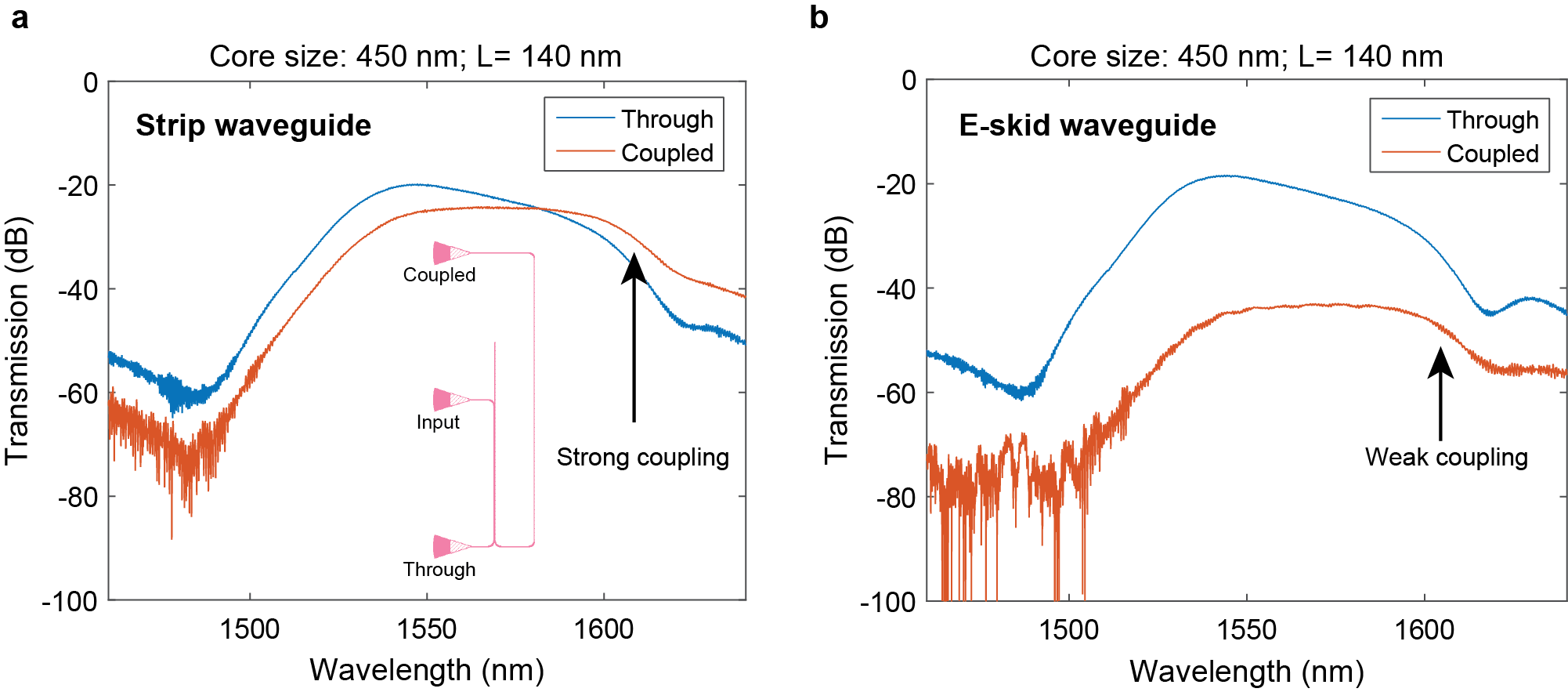}

\end{tabular}
\caption{{\bf Measured transmission spectra to the through and coupled waveguides.} {\bf a,} Strip waveguide; {\bf b,} e-skid waveguide. (Inset) Schematic representation of the experimental setup to measure the coupling length. Light is in-coupled through the middle grating coupler. The waveguide core size and the coupling distance ($L$) is 450 nm and 140 nm,
respectively, in both cases. The other parameters of the waveguides are the same as the waveguides in Fig. 4 in the main text. The anisotropic cladding causes coupling to the second waveguide to drop almost 20 dB. This helps to reduce the cross-talk in dense photonic integrated circuits.}
\label{fig:FigS_trans_measurement}
\end{figure*}

In the main text, we have reported the coupled power at a single wavelength. However, the coupling between two waveguides is a function of frequency in general. Supplementary Figure~\ref{fig:FigS_trans_measurement} displays the spectral response of strip and e-skid couplers. Since multilayer metamaterials are non-resonant broadband structures, multilayers show strong anisotropy for a wide range of wavelengths as long as the period is considerably smaller than the wavelength. Hence, as shown in Supplementary Fig.~\ref{fig:FigS_trans_measurement}, the multilayer cladding causes the cross-talk to decrease for a broad range of frequencies. 

\begin{figure}[ht]
\centering
\begin{tabular}{cc}

\includegraphics{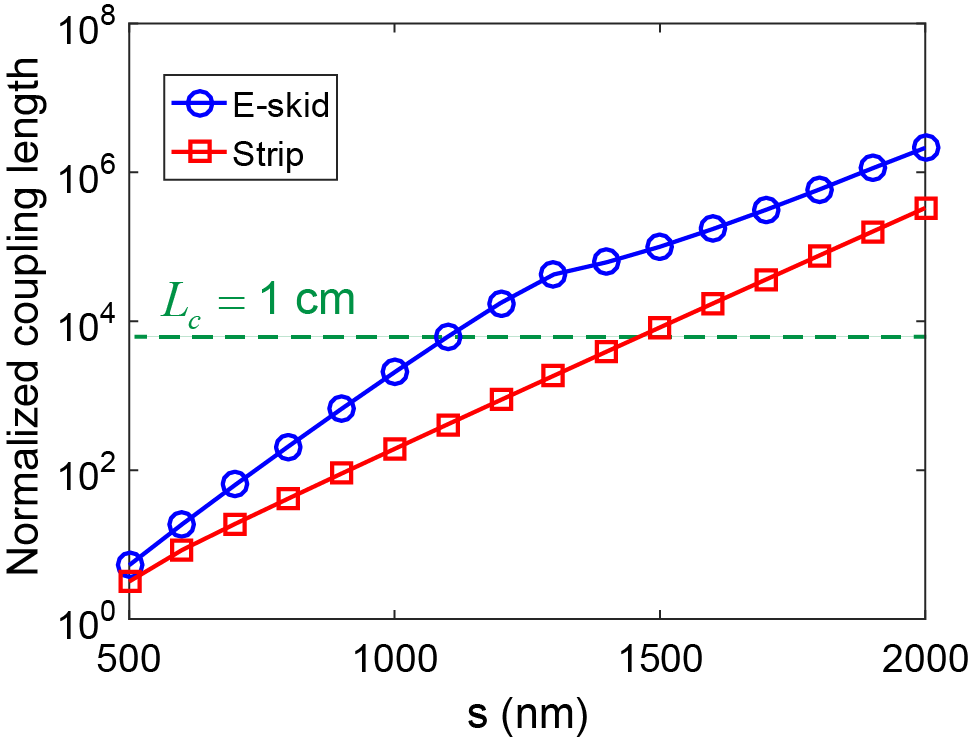}

\end{tabular}
\caption{{\bf Normalized coupling length ($L_c/\lambda$) versus the center-to-center separation between the coupled waveguides ($s$).} $w_0=450$~nm, $\Lambda=100$~nm, and $\rho=0.5$. Other simulation parameters are the same as those in Fig. 3 in the main text. $L_c=1$~cm has been defined as the minimum decoupling length. In this case, the minimal separation length (center-to-center) in e-skid waveguides and strip waveguides are 1104 nm and 1466 nm, respectively. This shows that e-skid waveguides will have higher performance for dense photonic integration than strip waveguides.}
\label{fig:FigS_separation}
\end{figure}

An important challenge of alternative waveguides is mode conversion efficiency from the grating coupler to the waveguide. The comparison of the output level of the through waveguide shows that the coupling efficiency from the grating coupler to the e-skid waveguides is very close to the coupling efficiency in strip waveguides. Moreover, as we explain in Supplementary note 5, the mode conversion efficiency from a strip waveguide to an e-skid waveguide with the same core size is above 98\%. This means that e-skid waveguides can easily be integrated to the silicon photonics chip.
Supplementary Figure~\ref{fig:FigS_separation} illustrates the normalized coupling length versus the center-to-center separation for coupled e-skid and strip waveguides. For a compact photonic circuit,  minimum separation with maximum decoupling between adjacent waveguides are required \cite{dai_comparative_2007}. However, as it is seen in Supplementary Fig.~\ref{fig:FigS_separation}, the coupling length exponentially decreases as the separation is reduced. This trade-off limits the integration density of photonic circuits. Due to the reduced skin-depth in e-skid waveguides, we can reduce the decoupled separation further and increase the integration density of photonic integrated circuits. We should emphasize that the decoupled separation in slot waveguides and photonic crystal waveguides are even more than that in strip waveguides \cite{dai_comparative_2007}.

\begin{figure*}[ht]
\centering
\begin{tabular}{cc}
\includegraphics{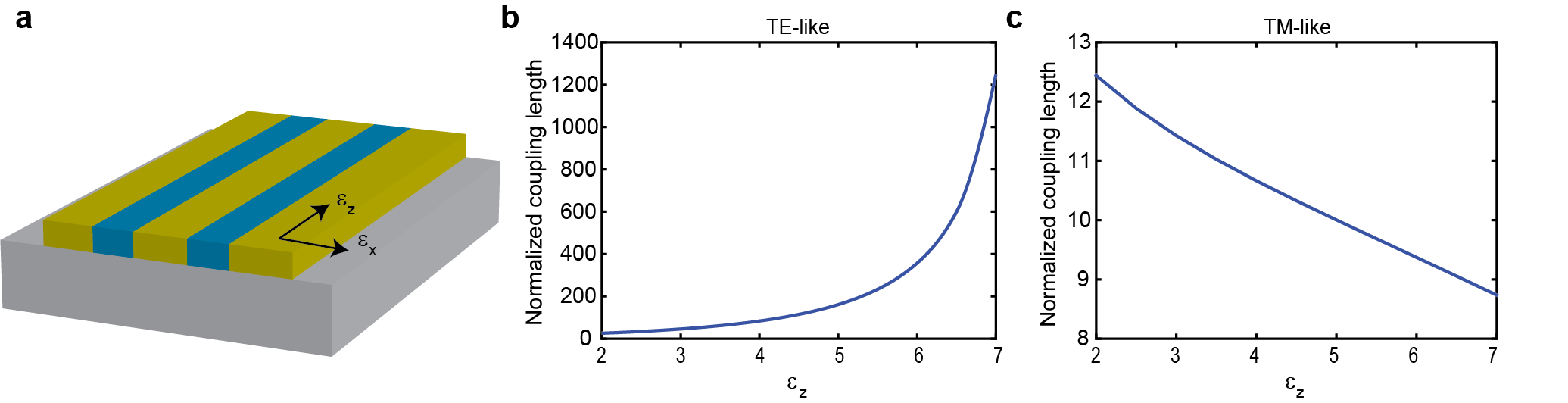}

\end{tabular}
\caption{{\bf Effect of cladding anisotropy on the coupling length for TE-like and TM-like modes of e-skid waveguides.} {\bf a,} Schematic representation of the two coupled Si waveguides with a core size of 300 nm, height of 220 nm, and center-to-center separation of 1000 nm operating at $\lambda=1550$ nm. The waveguides have been covered by a homogeneous anisotropic metamaterial (yellow) with $\varepsilon_x=2$ and $\varepsilon_y=\varepsilon_z$. {\bf b,} Coupling length for TE-like mode versus $\varepsilon_z$. {\bf c,} Coupling length for TM-like mode versus $\varepsilon_z$. As we enhance the anisotropy of the cladding, the coupling length of TE-like mode increases as a result of skin-depth reduction in the cladding.}
\label{fig:FigS_CT_TE_TM}
\end{figure*}

We also contrast the effect of anisotropy on TE-like and TM-like modes of on-chip e-skid waveguides. Supplementary Figure~\ref{fig:FigS_CT_TE_TM} shows the coupling length for TE-like and TM-like modes versus $\varepsilon_z$ while $\varepsilon_x$ is fixed in the cladding. As the anisotropy increases, we can reduce the skin-depth in the cladding for TE-like mode. Thus, the coupling length between the two coupled waveguides is strongly enhanced because the overlap  between evanescent waves of adjacent waveguides is suppressed. On the other hand, since the electric field in the $x$ direction is negligible for TM-like modes, these modes cannot feel the anisotropy of the cladding. Hence, the anisotropic cladding cannot help to control the momentum of evanescent waves for TM-like modes. 

\begin{figure*}[ht]
\centering
\begin{tabular}{cc}
\includegraphics{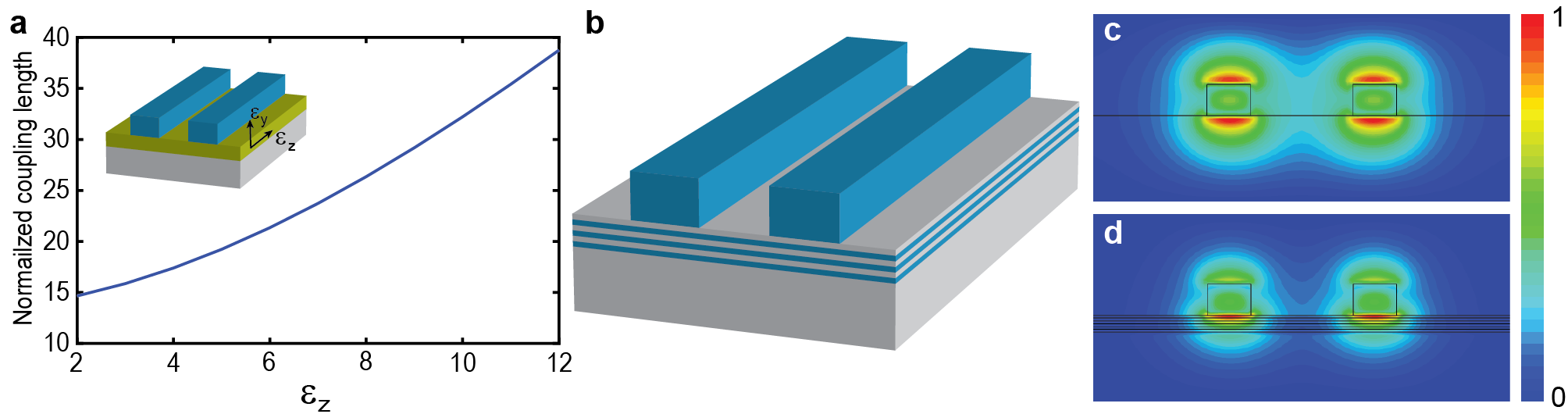}

\end{tabular}
\caption{{\bf E-skid waveguides for controlling the evanescent waves of the TM-like mode.} {\bf a,} The schematic representation of proposed e-skid waveguides for controlling the skin-depth of the TM-like mode (inset). Since the electric field of the TM-like mode is polarized in the $y$ direction, the cladding must be anisotropic in $yz$ plane. To simplify the structure, the anisotropic cladding (yellow) is implemented only underneath the waveguides. The plot shows the normalized coupling length between the two waveguides versus the anisotropy of the metamaterial layer while the permittivity in the $y$ direction is fixed to be 2.2. The thickness of the layer is 120 nm. The waveguide parameters are the same as that in Supplementary Fig.~\ref{fig:FigS_CT_TE_TM}. {\bf b,} Practical Si/SiO$_2$ multilayer structure to confine the evanescent waves and reduce the cross-talk between the two waveguides. $\Lambda=40$~nm and the silicon filling fraction is $\rho=0.5$. {\bf c} and {\bf d,} The normalized electric field profile of the even mode of the coupled strip and e-skid waveguides, respectively. It is seen that the multilayer structure helps to confine the mode inside the waveguide and reduces the overlap between the evanescent waves of the two waveguides. This results in the increase of the coupling length from $14.5\lambda$ to $30.6\lambda$.  
}
\label{fig:FigS_CT_TE_ML}
\end{figure*}

To control the evanescent waves of TM-like modes, the cladding must be anisotropic in the $yz$ plane and it must be added beneath or above the waveguide core. Supplementary Figure \ref{fig:FigS_CT_TE_ML}a (inset) illustrates a simplified structure to control the skin-depth of TM-like modes.  We keep the permittivity in the $y$ direction fixed, but we change it in the $z$ direction to control the anisotropy. As we increase the permittivity in the $z$ direction, the optical mode is confined inside the core and the overlap between the evanescent tales is reduced. Hence, as shown in Supplementary Fig. \ref{fig:FigS_CT_TE_ML}a, the coupling length increases. Supplementary Figure \ref{fig:FigS_CT_TE_ML}b displays a practical implementation of the anisotropic cladding for the TM-like modes composed of Si/SiO$_2$ multilayer claddings. In this case, $\Lambda=40$~nm and the silicon filling fraction is $\rho=0.5$. Supplementary Figure \ref{fig:FigS_CT_TE_ML}c and \ref{fig:FigS_CT_TE_ML}d show the normalized electric field profile of the even mode of the coupled strip and e-skid waveguides, respectively. It is seen that the skin-depth of the evanescent waves in e-skid waveguides is reduced, and as a result, the coupling length has been increased from $14.5\lambda$ for strip waveguides case to  $30.6\lambda$ for e-skid waveguides case. Note that if we add the multilayer on top of the waveguide core as well, more confinement is achievable and the cross-talk can be reduced further. 

\begin{figure*}[!ht]
\begin{center}
{\includegraphics[width=1.0\textwidth]{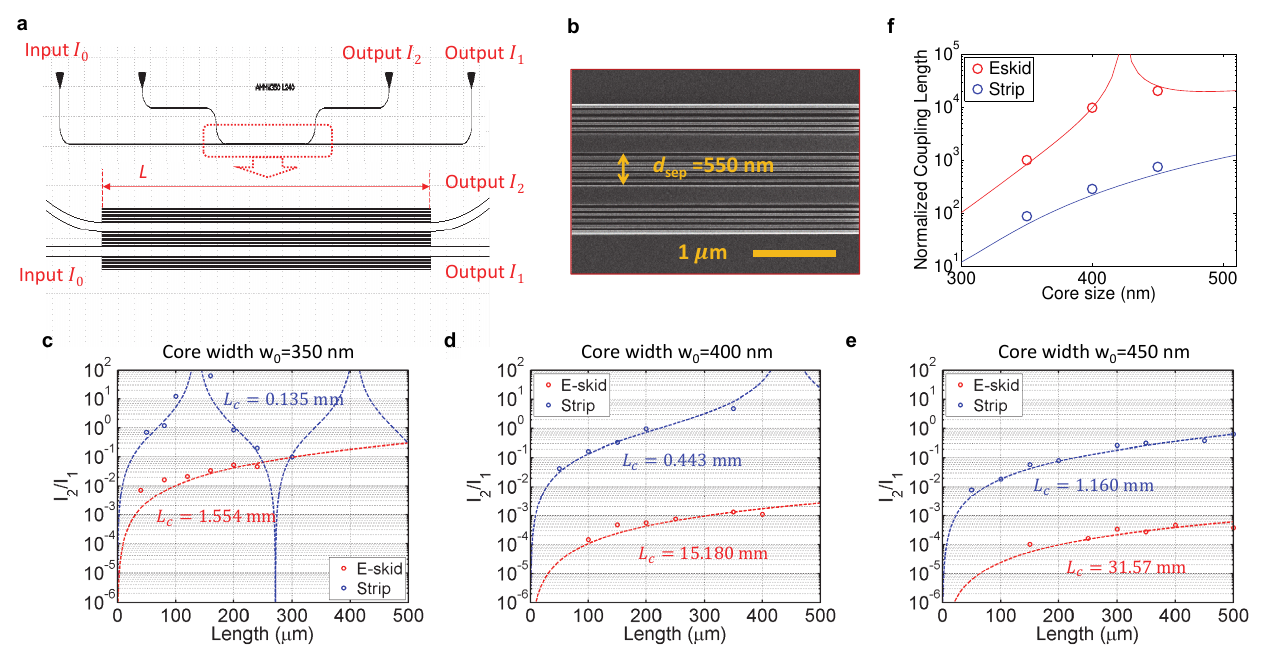}}
\caption{
{\bf Cross-talk in e-skid waveguides without an Upper cladding.}
{\bf a,} Schematic layout to measure the cross-talk and {\bf b,} SEM image of coupled e-skid waveguides on an SOI platform.
Geometric parameters are set to $h_0=220$~nm, $\Lambda=100$~nm, $\rho=0.5$, and $N=5$,
setting the separation distance between the two waveguides to be $d_{\rm sep}=550$~nm.
There is no upper cladding.
{\bf c-e,} Coupling length $L_{c}$ characterization by measuring the output power ratio ($I_2/I_1$) vs. device length $L$;
core widths $w_0$ are ({\bf c}) 350~nm, ({\bf d}) 400~nm,  and ({\bf e}) 450~nm, respectively.
Red and blue circles are the measured power ratio for the e-skid and strip waveguides, respectively,
and dashed lines are their fitting curves with $I_2/I_1=\tan^2(\pi L/2L_c)$.
{\bf f,} Normalized coupling length ($L_c/\lambda$) for the e-skid (red circles) and strip (blue circles) waveguides.
Red and blue lines are their respective simulation results.
}
\label{fig:FigS_CT_measurement}
\end{center}
\end{figure*}

\begin{figure*}[htbp]
\centering
\begin{tabular}{cc}

\includegraphics{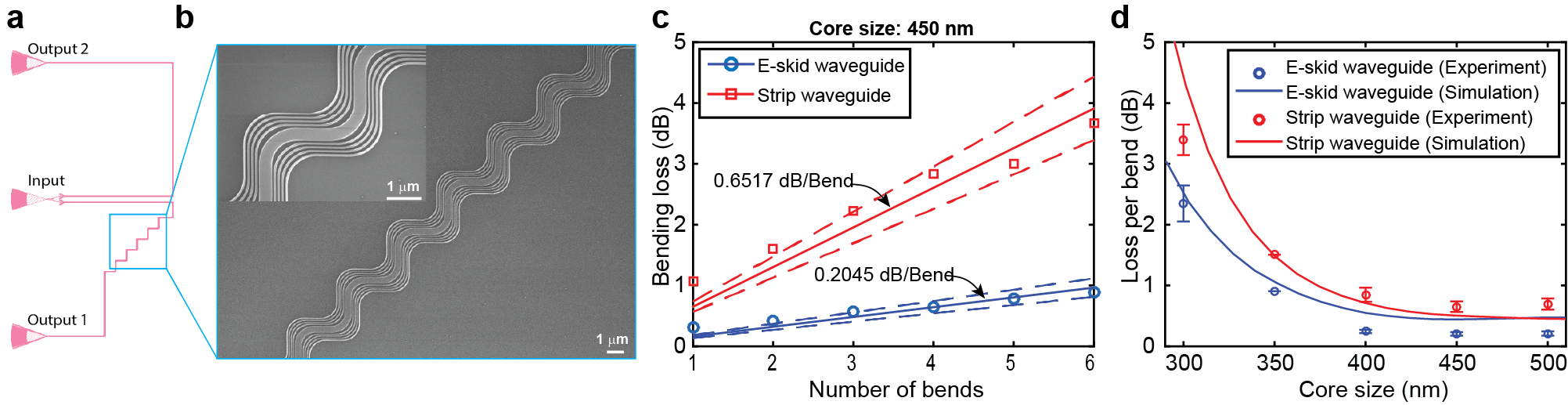}

\end{tabular}
\caption{{\bf Bending loss.} {\bf a,} Schematic representation of the experimental setup to measure the bending loss. Light is in-coupled through the middle grating coupler. It is divided into two branches of equal length but with a different number of bends. {\bf b,} The top view SEM image of an e-skid waveguide with multiple bends. The inset shows a closer view. The bending radius is 1 $\mu$m and the other parameters of the waveguides are the same as the waveguides in Fig. 4 in the main text. {\bf c,} The comparison of bending loss between e-skid and strip waveguides at telecommunication wavelength. The experimental points are fitted by a straight line. {\bf d,} The comparison of simulated and measured bending loss in e-skid and strip waveguides versus the core size indicating a considerable reduction in the bending loss, specially when the core size is small.}
\label{fig:FigS_BL}
\end{figure*}

\begin{figure*}[htbp]
\centering
\begin{tabular}{cc}

\includegraphics{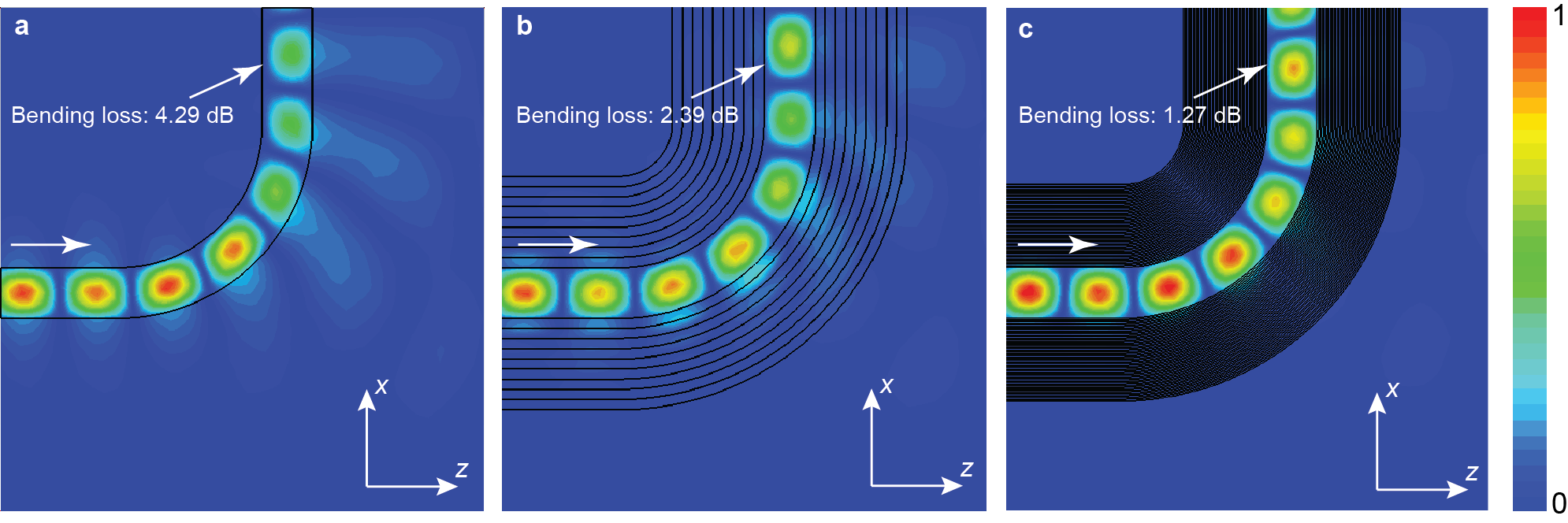}

\end{tabular}
\caption{{\bf Full-wave simulation of the TE-like mode of curved waveguides.} Full-wave simulation results of the magnetic field of the TE-like mode of curved ({\bf a}) strip waveguide, ({\bf b}) e-skid waveguide with $\Lambda$=120 nm, and ({\bf c}) e-skid waveguide with $\Lambda$=20 nm. The waveguide core size and bending radius are 300 nm and 1 $\mu$m, respectively. Other simulation parameters of the waveguides are the same as those in Supplementary Fig.~\ref{fig:FigS_BL}. The waveguides are excited from the left side. The anisotropic cladding reduces the skin-depth in the cladding. Thus, the scattering at the bend is reduced in the e-skid waveguide case. The comparison of the field amplitude at the end of the waveguides clearly shows that the bending loss in e-skid waveguide is relatively lower. In the ideal case where the periodicity is deep subwavelength, the bending loss is considerably lower in e-skid waveguides.}
\label{fig:FigS_BL_CST}
\end{figure*}

\begin{figure*}[htbp]
\centering
\begin{tabular}{cc}

\includegraphics[width=15cm]{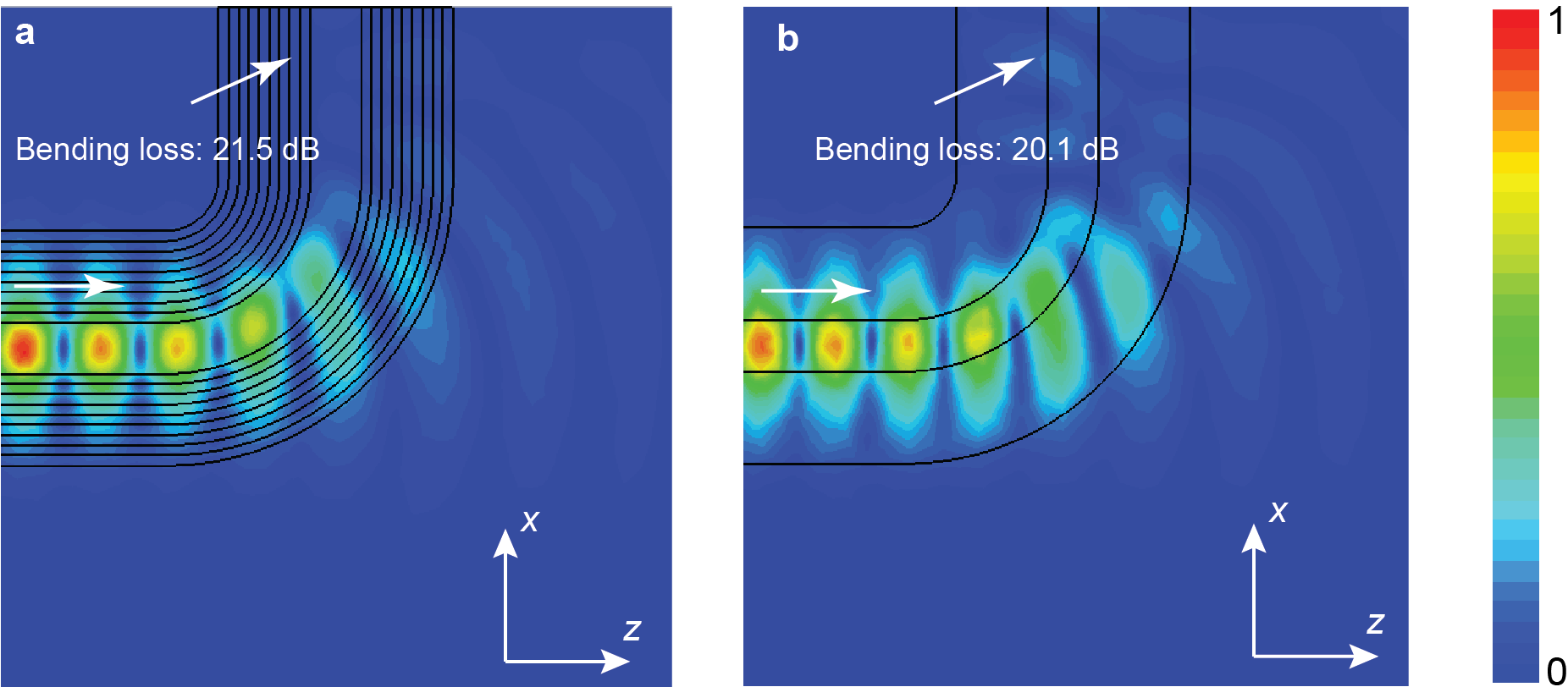}

\end{tabular}
\caption{{\bf Full-wave simulation of the TM-mode of curved waveguides.} {\bf a}, Full-wave simulation results of the magnetic field of the TM-like mode of curved e-skid waveguide with $\Lambda$=120 nm. Other simulation parameters of the waveguides are the same as those in Supplementary Fig.~\ref{fig:FigS_BL_CST}. {\bf b}, The field profile of the same waveguide when the multilayer cladding is replaced by an isotropic cladding with the same permittivity as the effective permittivity of the multilayer metamaterial in the $y$ direction. Since the electric field of the TM-like mode is polarized in the $y$ direction and the electric field in the $x$ direction is negligible, the anisotropy does not play any role in confinement. Thus, the TM-like mode is similar to that of an isotropic cladding waveguide. Note the TE-like mode probes the anisotropy of the multi-layer cladding and is fundamentally different.}
\label{fig:FigS_BL_CST_TM}
\end{figure*}

\begin{figure*}[htbp]
\centering
\begin{tabular}{cc}

\includegraphics{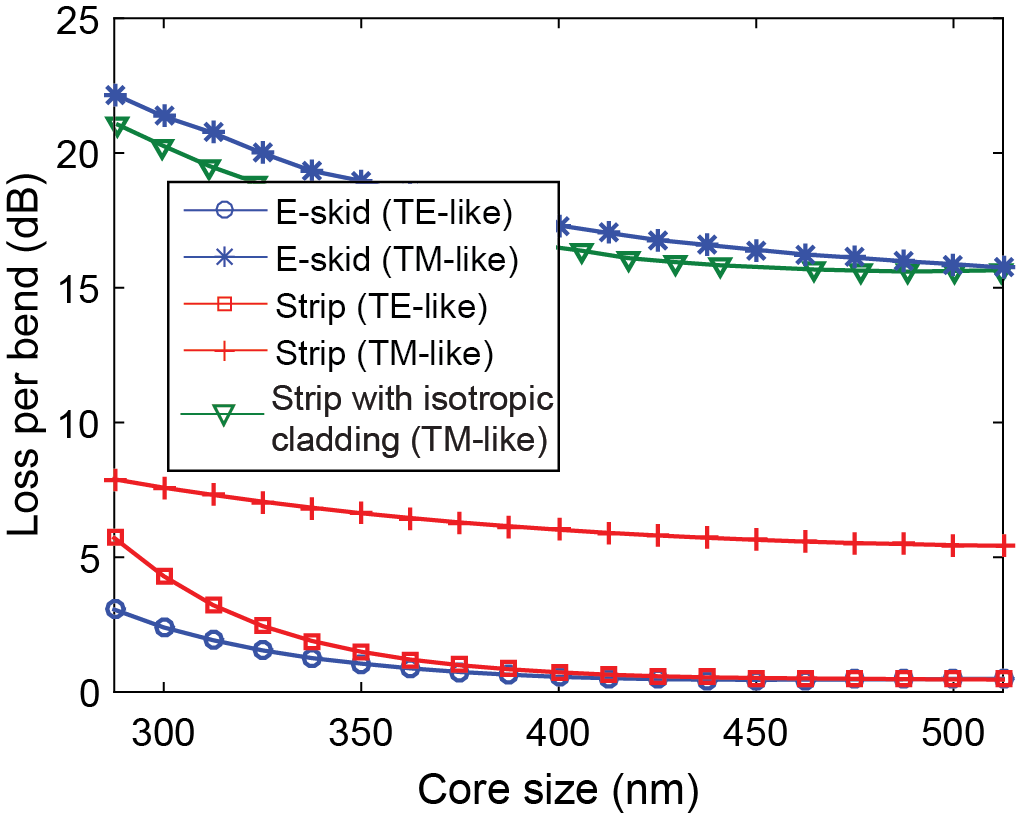}

\end{tabular}
\caption{{\bf Comparison of the full-wave simulated bending loss in e-skid and strip waveguides for both TE-like and TM-like modes.} The structure is the same as in Supplementary Fig.~\ref{fig:FigS_BL_CST}. The bending loss for TE-like modes in e-skid waveguide is lower in comparison with that in strip waveguides because the multilayer cladding confines the light inside the core and reduces the skin depth in the cladding, especially when the core size is small and a considerable fraction of the total power is in the cladding. The green curve shows the bending loss for the TM-like mode when the metamaterial cladding is replaced by an isotropic cladding with the same size and the same permittivity as the effective permittivity of the multilayer metamaterial in the $y$ direction. It is seen that the TM-like modes do not feel the anisotropy of the multilayer cladding. Thus, the anisotropic cladding cannot help to control the evanescent waves for TM-like modes. As a result, the bending loss for TM-like modes in e-skid waveguide is even worse than that in strip waveguides.}
\label{fig:FigS_BL_TE_TM}
\end{figure*}

\begin{figure*}[htbp]
\centering
\begin{tabular}{cc}
\includegraphics{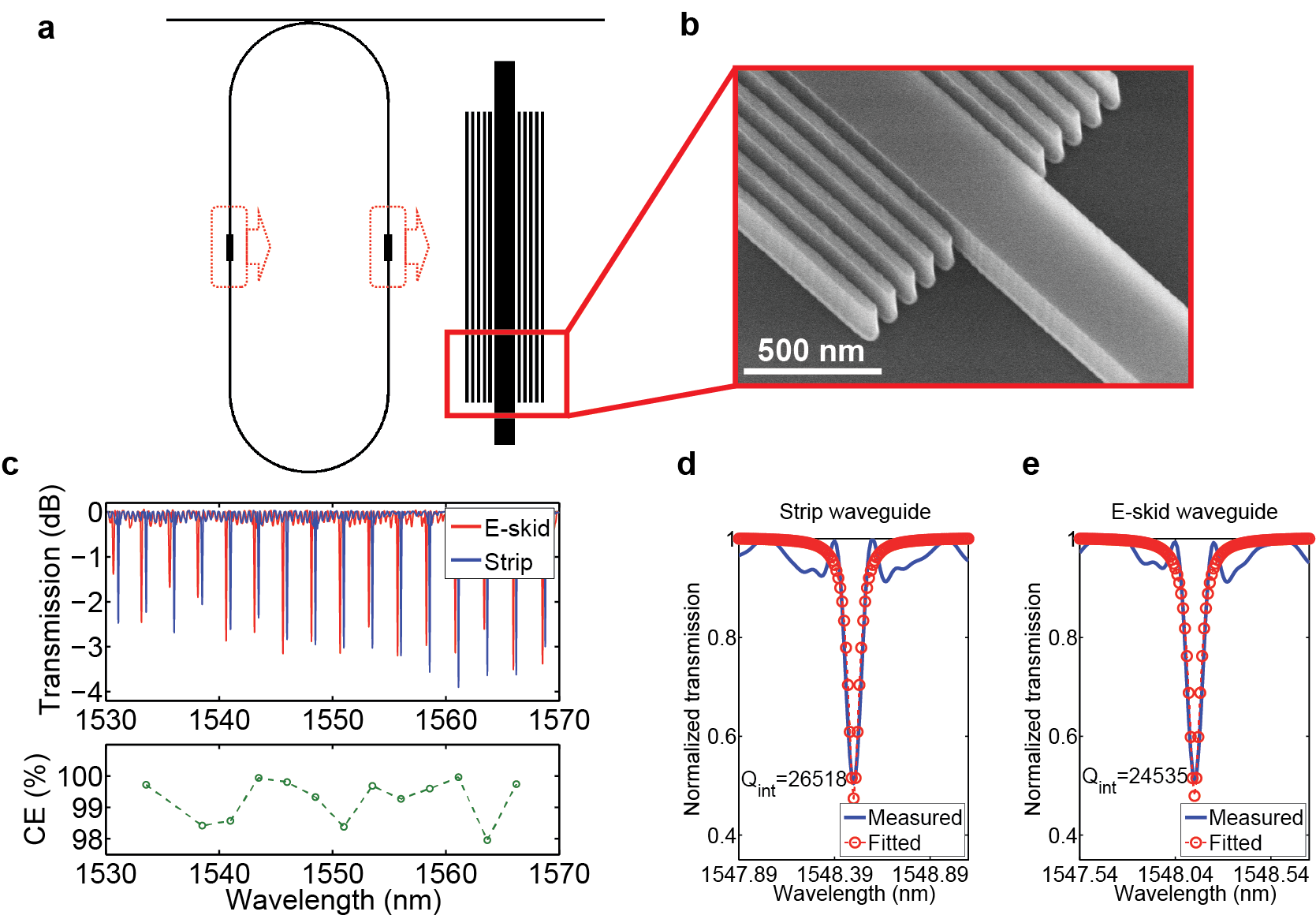}

\end{tabular}
\caption{{\bf Mode conversion efficiency.} {\bf a,} The schematic representation of the experiment set-up. {\bf b,} SEM image of an e-skid to strip waveguide transition. {\bf c,} Measured transmission spectra of the racetrack resonators without (blue) and with (red) the metamaterial claddings, and characterized mode conversion efficiency (CE). {\bf d} \& {\bf e,} Fitted resonances: ({\bf d}) without (strip waveguide) and ({\bf e}), with (e-skid waveguide) the metamaterial claddings. The waveguide geometries are set to $h_0 = 220$ nm, $w_0 = 350$ nm, $w\rho= w(1-\rho)= 50$ nm, and $N =$ 5.}
\label{fig:FigS_IL}
\end{figure*}

\subsection*{\label{sec:level2}Cross-talk in E-skid Waveguides Without an Upper Cladding}

{
In an e-skid waveguide, the waveguide cross-talk can be reduced further by increasing the anisotropy of the metamaterial.
Supplementary Figure~\ref{fig:FigS_CT_measurement}a shows the schematic layout to measure the cross-talk between coupled e-skid waveguides, and Supplementary Fig.~\ref{fig:FigS_CT_measurement}b shows an SEM image of coupled e-skid waveguides.
The parameters for the e-skid waveguide are set to $h_0=220$~nm, $\Lambda=100$~nm, $\rho=0.5$, and $N=5$;
this sets the separation distance between the two waveguides (edge-to-edge) to be $d_{\rm sep}=550$~nm, which is less than a free space wavelength ($\lambda=1550$~nm).
In this case, there is no upper SiO$_2$ cladding and the cross-section is similar to that in Fig.~3c of the main manuscript (Si/air multilayers).
This leads to a higher anisotropy in metamaterials and reduces the skin-depth of the guided mode compared to the case of Si/SiO$_2$ multilayers.
To evaluate the waveguide cross-talk, as in the main text,
we have characterized the coupling length $L_c$ by measuring the output power ratio ($I_2/I_1$) for different lengths ($L$) of devices.
Red and blue circles in Supplementary Figs.~\ref{fig:FigS_CT_measurement}c-\ref{fig:FigS_CT_measurement}e are the measured output power ratio for e-skid and strip waveguides, respectively,
and dashed lines are their respective fitting curves with $I_2/I_1=\tan^2(\pi L/2L_c)$.
Supplementary Figures~\ref{fig:FigS_CT_measurement}c-\ref{fig:FigS_CT_measurement}e are with different core widths of $w_0=$350, 400, and 450~nm, respectively.
Supplementary Figure~\ref{fig:FigS_CT_measurement}f summarizes the normalized coupling lengths ($L_c/\lambda$) that are characterized through Supplementary Figs.~\ref{fig:FigS_CT_measurement}c-\ref{fig:FigS_CT_measurement}e:
e-skid (red circles) and strip (blue circles) waveguides.
The solid lines are their respective simulation results that match well with the experimental measurements.
Notice that, in every case, the coupling length of an e-skid waveguide is much longer than that of a strip waveguide.
The maximum coupling length of more than 3~cm is achieved ($w_0=450$~nm),
and, when $w_0$ is 400 or 450~nm, the coupling length for an e-skid waveguide is about 30 times longer in comparison with strip waveguide.

\begin{figure*}[ht]
\centering
\begin{tabular}{cc}

\includegraphics{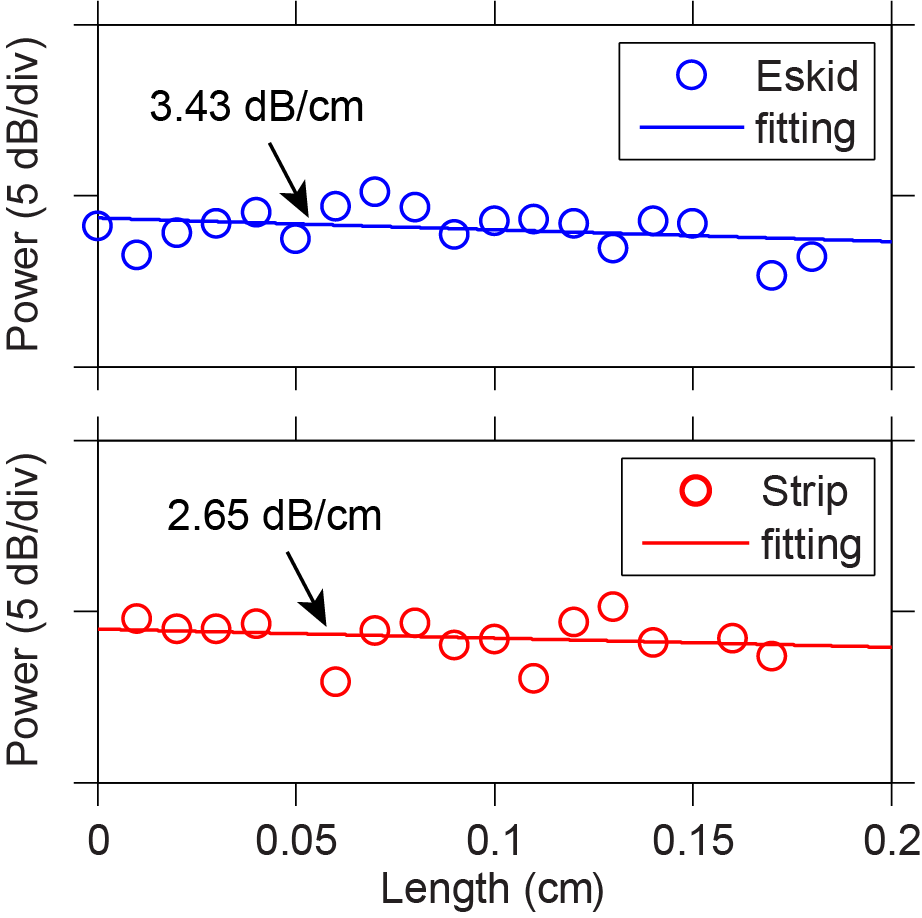}

\end{tabular}
\caption{{\bf Propagation losses in e-skid and strip waveguides.}  The circles represent normalized transmission through straight e-skid (blue) and strip (red) waveguides at $\lambda=1550$~nm. The length of the waveguides is varied from 0 to 1.8~mm to characterize the propagations loss (fitting curves). The propagation loss in e-skid and strip waveguide is 3.43 and 2.65 dB/cm at $\lambda=1550$~nm, and the average losses for e-skid and strip waveguides for wavelengths between 1540~nm to 1560~nm are 3.67 dB/cm and 1.84 dB/cm, respectively, with a standard deviation of 1.0 dB/cm and 1.4 dB/cm, respectively. The other waveguide parameters are $w_0=450$~nm, $h_0=220$~nm, $w_{\rho}=w_{(1-\rho)}=50$~nm, and $N=5$. This clearly shows that e-skid waveguides have comparable losses to standard strip waveguides opening the path for practical applications.}
\label{fig:FigS_PL}
\end{figure*}

\section{\label{sec:level1}Bending loss}

We have also simulated \cite{CST} and measured the bending loss when the waveguides are coated by an SiO$_2$ upper cladding (Supplementary Fig.~\ref{fig:FigS_BL}). An oxide cladding is usually used to protect the device and to integrate it with electronic interconnects \cite{chrostowski_silicon_2015}. In this experiment, the input power is divided into two paths with equal lengths. It is seen that the bending loss is considerably reduced due to the metamaterial cladding even when the core size is large.

In the main text, we have presented the magnetic field of curved strip waveguides and e-skid waveguides using a transformation optics approach to demonstrate the effect of skin-depth on the bending loss. However, for accurate calculation of bending loss, 3D full-wave simulations are required. Supplementary Figure~\ref{fig:FigS_BL_CST} shows the top view of  the simulated \cite{CST} magnetic field profile of TE-like mode for curved strip and e-skid waveguides. It is seen that for all-cases the skin-depth extends on the right side and is compressed on the left side in agreement with the transformation optics calculations. The performance of e-skid waveguides improves if we approach the effective medium theory limit ($\Lambda \rightarrow 0$) (Supplementary Fig.~\ref{fig:FigS_BL_CST}). 
As we mentioned earlier, the TM-like mode of on-chip waveguides does not feel the anisotropy of the cladding because the electric field in the $x$ direction is negligible. Hence, the confinement of the TM-like mode in e-skid cladding is the same as that in an isotropic cladding as shown in Supplementary Fig.~\ref{fig:FigS_BL_CST_TM}.

We compare the bending loss for both TE-like and TM-like modes of e-skid and strip waveguides in Supplementary Fig.~\ref{fig:FigS_BL_TE_TM}. As we mentioned in the main text, the additional degree of freedom in total internal reflection can be used to control evanescent waves of p-polarized (TE-like) modes. Hence, the anisotropic cladding helps to reduce the bending loss of the TE-like mode in e-skid waveguides in comparison with strip waveguides. However, this degree of freedom does not exist for s polarized (TM-like) modes. Since the effective permittivity of the cladding for the TM-like mode is larger than the permittivity of SiO$_2$, the evanescent waves decay slower for TM-like modes in e-skid waveguides. Thus, the skin-depth extends more and more power is radiated at sharp bends in comparison with strip waveguides.

\section{\label{sec:level1}Mode conversion efficiency}

The mode conversion efficiency between the e-skid waveguide and the strip waveguide are also investigated. For the mode conversion efficiency, we use a racetrack resonator (Supplementary Fig.~\ref{fig:FigS_IL}) to characterize the insertion loss efficiency \cite{han_strip-slot_2016}. Supplementary Figure~\ref{fig:FigS_IL}c shows the normalized transmission spectra for the e-skid (red) and strip (blue) waveguides. The resonances are slightly shifted with metamaterials due to the different propagation phase in the e-skid waveguide (see Fig. 3 in the main text), but the extinction ratios are almost similar suggesting a similar cavity loss. Supplementary Figures~\ref{fig:FigS_IL}d and \ref{fig:FigS_IL}e  are the fitted resonances for the strip and e-skid waveguides, respectively, and the intrinsic quality factors are 26,518 and 24,535, which correspond to the round-trip losses of 0.6893 dB/round and 0.7466 dB/round, respectively \cite{xiao_modeling_2007}. Since we have four interfaces for the mode conversion, a rough estimation of the mode conversion loss is about 0.0143 dB/facet, which corresponds to a mode conversion efficiency of 99.6\%. The mode conversion efficiencies (CEs) with other resonances are also characterized and plotted in Supplementary Fig.~\ref{fig:FigS_IL}c. This high mode conversion efficiency between the strip waveguide and the e-skid waveguide in TE-like modes indicates a low insertion loss, and suggests a high compatibility with the previously used PIC devices.
\\

\section{\label{sec:level1}Propagation loss}
To characterize the propagation losses in the e-skid and the strip waveguides, we have fabricated two sets of waveguides (e-skid and strip waveguides) with different lengths of the straight section. The waveguide parameters are set to $w_0=450$~nm, $h_0=220$~nm, $w_{\rho}=w_{(1-\rho)}=50$~nm, and $N=5$; then the lengths of the straight section of the e-skid and strip waveguides are varied from 0 to 1.8~mm. Supplementary Figure~\ref{fig:FigS_PL} shows the measured transmission powers of e-skid (blue circles) and strip (red circles) waveguides at $\lambda_0=1550$~nm, for different device lengths. Solid lines in each figure are the linear fitting curves that give the propagation losses of 2.65 dB/cm and 3.43 dB/cm to strip and e-skid waveguides, respectively. The propagation losses for strip and e-skid waveguides at different wavelengths have also been characterized. The average losses for strip and e-skid waveguides for wavelengths between 1540~nm to 1560~nm are 1.84 dB/cm and 3.67 dB/cm, respectively, with a standard deviation of 1.4 dB/cm and 1.0 dB/cm, respectively. This propagation loss of the e-skid waveguide is reasonable for compact devices, especially given that the cross-talks and bending losses are improved significantly (see Table I in the main text for comparison with other techniques).     
\clearpage
\bibliography{main}
\end{document}